\documentclass[pra,aps,10pt,superscriptaddress,twocolumn,floatfix]{revtex4-1}  

\usepackage{graphicx,color}
\usepackage{amsmath,amssymb,bm}
\usepackage{braket}		

\DeclareMathOperator{\Tr}{Tr}

\usepackage[plainpages=false,pdfpagelabels,colorlinks=true,linkcolor=red,urlcolor=blue,citecolor=blue,pdftitle={Titl}e,pdfauthor={},pdfdisplaydoctitle=true,pdfduplex=DuplexFlipLongEdge]{hyperref}

\definecolor{darkred}{rgb}{0.90,0.2,0.2}
\definecolor{darkgreen}{rgb}{0,0.60,.2}
\definecolor{darkblue}{rgb}{0.1,0.3,1}
\definecolor{grey}{cmyk}{0,0,0,0.25}
\definecolor{orange}{cmyk}{0,0.6,0.8,0}

\begin{document}
\title{Statistical properties of the off-diagonal matrix elements of observables\\ in eigenstates of integrable systems}

\author{Yicheng Zhang}
\affiliation{Department of Physics, The Pennsylvania State University, University Park, Pennsylvania 16802, USA}
\author{Lev Vidmar}
\affiliation{Department of Theoretical Physics, J. Stefan Institute, SI-1000 Ljubljana, Slovenia}
\affiliation{Department of Physics, Faculty of Mathematics and Physics, University of Ljubljana, SI-1000 Ljubljana, Slovenia}
\author{Marcos Rigol}
\affiliation{Department of Physics, The Pennsylvania State University, University Park, Pennsylvania 16802, USA}

\begin{abstract}
We study the statistical properties of the off-diagonal matrix elements of observables in the energy eigenstates of integrable quantum systems. They have been found to be dense in the spin-1/2 XXZ chain, while they are sparse in noninteracting systems. We focus on the quasimomentum occupation of hard-core bosons in one dimension, and show that the distributions of the off-diagonal matrix elements are well described by generalized Gamma distributions, in both the presence and absence of translational invariance but not in the presence of localization. We also show that the results obtained for the off-diagonal matrix elements of observables in the spin-1/2 XXZ model are well described by a generalized Gamma distribution.
\end{abstract}

\maketitle

\section{Introduction} \label{sec:intro}

Whether thermalization occurs in isolated quantum many-body systems has attracted much attention since the birth of quantum mechanics~\cite{Neumann_29}. A recent impetus for exploring this question comes from experimental advances in ultracold quantum gases, in which nearly isolated quantum systems are realized and used routinely as quantum simulators~\cite{Lewenstein_Sanpera_2007, Bloch_Dalibard_08, Bloch_Dalibard_2012, Langen_Geiger_15, eisert_friesdorf_review_15}. Thermalization has been observed experimentally in nonintegrable (quantum-chaotic interacting) systems~\cite{Trotzky2012, Kaufman2016, clos_porras_16, tang_kao_18}, as it had been observed earlier in numerical simulations~\cite{rigol_dunjko_08, Rigol_09_Breakdown, Rigol_09_Quantum} (see Ref.~\cite{dalessio_kafri_16} for a review). On the other hand, lack of thermalization has been observed experimentally in near-integrable systems~\cite{Kinoshita2006, gring_kuhnert_12, langen15a, wilson_malvania_20, Malvania_Zhang_21}, as well as in early numerical simulations of integrable quantum dynamics~\cite{rigol_dunjko_07, rigol_muramatsu_06} (see Ref.~\cite{vidmar16} for a review). Integrable systems are the focus of this work.

Thermalization in nonintegrable systems is understood in terms of the eigenstate thermalization hypothesis (ETH)~\cite{deutsch_91, srednicki_94, srednicki_99, rigol_dunjko_08, dalessio_kafri_16}. The ETH can be written as an ansatz for the matrix elements of few-body observables $O_{\alpha\beta}\equiv\langle\alpha|\hat O|\beta\rangle$ in the energy eigenstates $\{|\alpha\rangle\}$~\cite{srednicki_99, dalessio_kafri_16},
\begin{equation}\label{eq:ETH}
    O_{\alpha\beta}=O(\bar E)\delta_{\alpha\beta}+e^{-S(\bar E)/2} f_O (\bar E, \omega)R_{\alpha\beta}\,,
\end{equation}
where the average energy of pairs of eigenstates is $\bar E=(E_\alpha+E_\beta)/2$, the difference is $\omega=E_\alpha-E_\beta$, $S(\bar E)$ is the thermodynamic entropy at energy $\bar E$, $R_{\alpha\beta}$ is a random (in general, normally distributed) variable with zero mean and unit variance, and $O(\bar E)$ and $f_O (\bar E, \omega)$ are smooth functions of their arguments. Since the thermodynamic entropy is an extensive quantity away from the edges of the spectrum, $e^{-S(\bar E)/2}$ is exponentially small in the system size. For $\bar E$ close to the center of the energy spectrum, $e^{-S(\bar E)/2}\simeq 1/\sqrt{D}$, where $D$ is the size of Hilbert space. The smoothness of the diagonal matrix elements as functions of the energy $\bar E$ makes the agreement between the observable after equilibration and statistical mechanics possible, while the smallness of the off-diagonal matrix elements ensures the smallness of the temporal fluctuations after equilibration.

Thanks to many computational studies, over the last fifteen years we have sharpened our understanding of the differences between integrable systems (which do not exhibit eigenstate thermalization) and nonintegrable ones (which do), see, e.g., Refs.~\cite{rigol_dunjko_08, Rigol_09_Breakdown, Rigol_09_Quantum, Santos_Rigol_10, Biroli_Kollath_10, Khatami_Pupillo_13, Ikeda_Watanabe_13, Beugeling_Moessner_14, beugeling_moessner_15, Alba_15, LeBlond_Mallayya_19, Mierzejewski_Vidmar_20, LeBlond_Rigol_Eigenstate_20}, and Ref.~\cite{dalessio_kafri_16} for a review. Within integrable systems, we have also learned about the crucial effect of interactions, and that noninteracting systems are very special (as we will discuss later). The presence of interactions, even in models that can be mapped onto noninteracting ones (such as hard-core boson models), results in integrable dynamics that is fundamentally different from that in noninteracting systems~\cite{wright_rigol_14}. 

For a paradigmatic integrable interacting model, the spin-1/2 XXZ chain, two important observations have been made recently about the matrix elements of observables in energy eigenstates~\cite{LeBlond_Mallayya_19}. The first one is that the off-diagonal matrix elements are {\it dense} (the overwhelming majority does not vanish as it does in noninteracting systems, in which they are {\it sparse}). One can therefore define a meaningful function $V_O(\bar E,\omega)=e^{S(\bar E)}|\overline{O_{\alpha\beta}|^2}$, which we refer to as the scaled variance. It can be seen as the analog of the $|f_O(\bar E,\omega)|^2$ function in Eq.~\eqref{eq:ETH}. $V_O(\bar E,\omega)$ has been shown to be a smooth function of $\omega$, fixing $\bar E$ to be at the center of the spectrum, for various observables~\cite{LeBlond_Mallayya_19, LeBlond_Rigol_Eigenstate_20, Brenes_LeBlond_20, brenes_goold_20}. We note that $|f_O(\bar E,\omega)|^2$ for nonintegrable models, and $V_O(\bar E,\omega)$ for integrable ones, control (together with the initial state) the dynamics of the specific observable. Those functions can be probed experimentally, e.g., by measuring heating rates~\cite{Mallayya_Rigol_19}. The second observation is about the distribution of the off-diagonal matrix elements, and it is the focus of this work. In contrast to the Gaussian distributions of matrix elements that are generic for nonintegrable systems~\cite{beugeling_moessner_15, luitz_barlev_16, khaymovich_haque_19, LeBlond_Mallayya_19, Brenes_LeBlond_20, brenes_goold_20, LeBlond_Rigol_Eigenstate_20, santos_perezbernal_20, noh_21, brenes_pappalardi_21}, the distributions of matrix elements in the spin-1/2 XXZ chain were found to be close to skewed log-normal-like distributions~\cite{LeBlond_Mallayya_19, brenes_goold_20, LeBlond_Rigol_Eigenstate_20}.

The distributions of matrix elements of observables in the spin-1/2 XXZ chain were studied using full exact diagonalization in the presence of translational invariance in Refs.~\cite{LeBlond_Mallayya_19, LeBlond_Rigol_Eigenstate_20}, and for chains with open boundary conditions in Ref.~\cite{brenes_goold_20}. Because of the exponential increase in complexity of those calculations with the chain size, the largest chains studied had $L=26$ sites. This prevented an accurate characterization of the distributions and of their scaling with the chain size. The main focus of this work are models of hard-core bosons in one-dimensional lattices, i.e., bosons that exhibit an infinite on-site repulsion, with particle-number conservation and no inter-site interactions. Such models are mappable onto noninteracting spinless fermion models. Our goal is to use them to gain a more accurate understanding of the distributions of off-diagonal matrix elements of observables in integrable models in the presence of interactions, and of their scalings with the system size.

We study the occupation of quasimomentum modes (nonlocal one-body observables), which can be measured in experiments with ultracold quantum gases. We consider both the translationally invariant model as well as the Aubry-Andr\'e model. The dynamics of various observables in the latter model were studied in Ref.~\cite{rigol_fitzpatrick_11}, where equilibration to the predictions of a generalized Gibbs ensemble was shown to occur in the delocalized regime. The dynamics of the same observables in the noninteracting spinless fermion model were studied in Ref.~\cite{He_Santos_13}, along with the diagonal matrix elements of the occupation of the zero quasimomentum mode in the hard-core boson model. The latter study revealed the expected lack of compliance with the ETH due to the integrability of the model. 

Here we discuss the differences between the behavior of the off-diagonal matrix elements of observables in the hard-core boson model and in the noninteracting spinless fermion model to which the former can be mapped. We then show that, in the delocalized regime of the hard-core boson model, the distributions of off-diagonal matrix elements of the occupation of the quasimomentum modes are well described by generalized Gamma distributions~\cite{gamma_distribution}. We also show that results reported in Ref.~\cite{LeBlond_Rigol_Eigenstate_20} for the distribution of the off-diagonal matrix elements of a local observable in the spin-1/2 XXZ chain are well described by a generalized Gamma distribution, suggesting that such distributions are generic in integrable interacting models.

The paper is organized as follows. In Sec.~\ref{sec:general}, we discuss the general differences between the off-diagonal matrix elements of few-body observables in systems consisting of noninteracting spinless fermions (which are sparse) and of hard-core bosons (which, for nonlocal observables, need not be sparse). In Sec.~\ref{sec:HCBstranslation}, we study the properties of the off-diagonal matrix elements of the occupation of the zero quasimomentum mode of hard-core bosons in the presence of translational invariance. Sections~\ref{sec:fermionsAA} and~\ref{sec:HCBsAA} are devoted to studying the effect of breaking translational invariance, as well as of localization, in the context of the Aubry-Andr\'e model. In Sec.~\ref{sec:fermionsAA} we discuss results for noninteracting fermions, while in Sec.~\ref{sec:HCBsAA} we discuss results for the corresponding model of hard-core bosons. A discussion of the relevance of our results beyond hard-core boson models is presented in Sec.~\ref{sec:xxz}. Specifically, we show that a generalized Gamma distribution describes the distribution of the off-diagonal matrix elements of an observable studied in Ref.~\cite{LeBlond_Rigol_Eigenstate_20} in the integrable spin-1/2 XXZ chain. We summarize our results in Sec.~\ref{sec:summary}.

\section{Noninteracting spinless fermions vs hard-core bosons} \label{sec:general}

We begin with a general discussion of the properties of the matrix elements of observables in noninteracting spinless fermion models and in hard-core bosons models. Having the quasimomentum occupation in mind, we identify important differences between the off-diagonal matrix elements of nonlocal few-body observables in both models.

\subsection{General results for \\ noninteracting spinless fermions} \label{sec:fermionsgeneral}

Let us begin by discussing properties of the off-diagonal matrix elements of observables in a general model of noninteracting spinless fermions with particle number conservation in a lattice with $L$ sites. The Hamiltonian can be written as is
\begin{equation}\label{eq:Hsf}
\hat H^{\rm SF}=-\sum_{\substack{i,j=1\\i\neq j}}^{L}(A_{ij}\hat f^\dagger_i\hat f^{}_j+\text{H.c.})+\sum_{i=1}^LV_i\hat f^\dagger_i\hat f^{}_i\,,
\end{equation}
where $\hat f^\dagger_i$ ($\hat f^{}_i$) creates (annihilates) a spinless fermion at site $i$, $A_{ij}$ is the hopping amplitude between sites $i$ and $j$, and $V_i$ is the magnitude of a local potential at site $i$. All many-body energy eigenstates $|\alpha\rangle$ of $\hat H^{\rm SF}$, for $N$ fermions, can be written as Slater determinants
\begin{equation}\label{eq:slater}
|\alpha\rangle=\prod_{m=1}^{N}\hat c^{\dagger}_{\alpha_m}|0\rangle\,,
\end{equation}
where 
\begin{equation}\label{eq:crea}
\hat c^\dagger_{\alpha_m}=\sum_{i=1}^L d_{\alpha_m}^i \hat f^\dagger_i
\end{equation}
creates a spinless fermion with eigenenergy $E_{\alpha_m}$ (the coefficients $d_{\alpha_m}^i$ implement the change of basis).

We are interested in the off-diagonal matrix elements of particle-number conserving observables $\hat O$ between energy eigenstates $|\alpha\rangle$ and $|\beta\rangle$ that have the same number of particles, namely, on $O_{\alpha\beta}=\langle\alpha|\hat O|\beta\rangle$. Let us assume that $\hat O$ can be expressed using at most $M$ pairs of creation and annihilation operators (say, in the site basis), with $M\leq {\rm min}(N,L-N)$,
\begin{align}\label{eq:operator}
\hat O=&\sum_{i_1i'_1}\sigma_{i_1i'_1}\hat f^\dagger_{i_1}\hat f^{}_{i'_1}\nonumber\\
&+\sum_{i_1i_2i'_1j'_2}\sigma_{i_1i_2i'_1i'_2}\hat f^\dagger_{i_1}\hat f^\dagger_{i_2}\hat f^{}_{i'_1}\hat f^{}_{i'_2}+\cdots\nonumber\\
&+\sum_{i_1\cdots i_Mi_1'\cdots i_M'}\sigma_{i_1\cdots i_Mi_1'\cdots i_M'}\hat f^\dagger_{i_1}\cdots\hat f^\dagger_{i_M}\hat f^{}_{i_1'}\cdots\hat f_{i_M'}\,,
\end{align}
where $\sigma_{...}$ are constants. Then, a necessary criterion for $O_{\alpha\beta}$ to be nonzero is that the analog of Eq.~(\ref{eq:slater}) for $|\beta\rangle$ contains at most $M$ single-particle operators $\hat c^\dagger_{\beta_m}$ that are not contained among the $N$ operators $\hat c^\dagger_{\alpha_m}$ in $|\alpha\rangle$. This follows after noticing that one can rewrite Eq.~\eqref{eq:operator} in terms of the creation (annihilation) operators $\hat c^\dagger_m$ ($\hat c^{}_m$), and this does not change the form of $\hat O$ in terms of the new operators (only the coefficients change).

Using this, one can find an {\it upper bound} for the number of nonzero off-diagonal matrix elements $O_{\alpha\beta}$,
\begin{equation}\label{eq:Nnonzero}
\bar{N}_{\rm nonzero}= {L \choose {N}}\sum_{j=1}^{M} {N \choose {j}}{L-N \choose{j}}\,,
\end{equation}
where ${L \choose {N}}$ is the number of many-body energy eigenstates, and $\sum_{j=1}^{j'}{N \choose {j}}{L-N \choose{j}}$ bounds the number of nonzero matrix elements that the terms in $\hat O$ with up to $j'$ pairs of creation and annihilation operators can generate for any given many-body energy eigenstate. Comparing $\bar{N}_{\rm nonzero}$ to the total number of $O_{\alpha\beta}$, which is $N_{\rm tot}= {L \choose {N}}\big[{L \choose {N}}-1\big]$, the fraction of nonzero off-diagonal matrix elements must be smaller than or equal to
\begin{equation}\label{eq:rnonzero}
r_{\rm nonzero}=\frac{\sum_{j=1}^{M} {N \choose {j}}{L-N \choose{j}}}{{L \choose {N}}-1}\,.
\end{equation}
Taking the thermodynamic limit, $N\rightarrow\infty$ and $L\rightarrow\infty$ with $N/L={\rm const}$ and a fixed $M$, results in a vanishing $r_{\rm nonzero}$. One usually refers to the operators $\hat O$ in Eq.~\eqref{eq:operator} as nonlocal few-body operators when $M$ is $O(1)$, namely, when $M$ is independent of $N$ and $L$.

In this work, we focus on the occupation of quasimomentum modes
\begin{equation}\label{eq:mk}
\hat{\mathsf{m}}_k=\frac{1}{L}\sum_{j,l=1}^{L}e^{ik(j-l)}\hat f^\dagger_j \hat f^{}_l\,,
\end{equation}
which can be considered as a special case of Eq.~(\ref{eq:operator}) with $M=1$. $\hat{\mathsf{m}}_k$ is a nonlocal one-body operator, and it can be measured in experiments with ultracold quantum gases in optical lattices~\cite{Bloch_Dalibard_08}. For a system with $L$ sites and $N$ particles, the square of the (properly normalized) Hilbert-Schmidt norm of $\hat{\mathsf{m}}_k$ is
\begin{equation}\label{eq:hsnorm}
||\hat{\mathsf{m}}_k||^2\equiv\frac{1}{D}\Tr\{\hat{\mathsf{m}}_k^2\} = \frac{1}{D} \sum_{\alpha,\beta=1}^D |\langle\alpha|\hat{\mathsf{m}}_k|\beta\rangle|^2 = \frac{N}{L}\,,
\end{equation} 
where $D = {L\choose N}$ is the size of the Hilbert space, at a given $N$ and $L$, over which the trace is computed.

It follows from Eq.~(\ref{eq:rnonzero}) that for $M=1$ the fraction of nonzero matrix elements is
\begin{equation} \label{def_rm0}
    r_{\mathsf{m}_0} = \frac{N (L-N)}{(D-1)} = \frac{L^2}{D}\frac{ n (1-n)}{(1-1/D)}\;,
\end{equation}
where we introduced the ``filling'' $n=N/L$.

For an average number of nonzero off-diagonal matrix elements $DN(L-N)$ of $(\mathsf{m}_0)_{\alpha\beta}$, as per Eq.~\eqref{eq:Nnonzero}, we can use the Hilbert-Schmidt norm from Eq.~\eqref{eq:hsnorm} to estimate their typical magnitude (assuming that all matrix elements are similar in magnitude). One gets that the typical nonzero matrix elements scale as
\begin{equation} \label{m0_typical}
    |(\mathsf{m}_0)_{\alpha\beta}|^2 \approx \frac{1}{L(L-N)} = \frac{1}{L^2} \frac{1}{(1-n)}\;.
\end{equation}

Summarizing our discussion so far, the fraction of nonzero off-diagonal matrix elements of few-body operators in many-body energy eigenstates of models of noninteracting spinless fermions vanishes in the thermodynamic limit~\cite{Khatami_Pupillo_13, haque_mcclarty_19}. The specific results obtained here for $(\mathsf{m}_0)_{\alpha\beta}$ will be used in our discussion in Sec.~\ref{sec:fermionsAA}.

\subsection{General results for hard-core bosons} \label{sec:HCBsgeneral}

Next we turn our attention to the most general (particle-number conserving) model of hard-core bosons in one dimension that can be mapped onto a model of noninteracting spinless fermions. The Hamiltonian has the form
\begin{equation}\label{eq:Hhcb}
\hat H^{\rm HCB} = -\sum_{i=1}^{L}(A_{i,i+1}\hat b^\dagger_i\hat b^{}_{i+1}+\text{H.c.})+\sum_{i=1}^L V_i \hat b^\dagger_i\hat b_i\,,
\end{equation}
where $\hat b^\dagger_i$ ($\hat b_i$) creates (annihilates) a hard-core boson at site $i$, $A_{i,i+1}$ is the hopping amplitude between nearest-neighbor sites $i$ and $i+1$, and $V_i$ is the magnitude of a local potential at site $i$. Periodic boundary conditions are assumed in Eq.~(\ref{eq:Hhcb}), i.e., $\hat b_{L+1} \equiv \hat b_1$ and $A_{L,L+1} \equiv A_{L,1}$. The hard-core constraint $\hat b^{\dagger2}_i=\hat b^{2}_i=0$ prevents  two (or more) bosons from occupying the same lattice site.

The hard-core boson Hamiltonian $\hat H^{\rm HCB}$ in Eq.~(\ref{eq:Hhcb}) can be mapped onto a similar Hamiltonian of noninteracting spinless fermion Hamiltonian, specifically, onto $\hat H^{\rm SF}$ in Eq.~(\ref{eq:Hsf}) in one dimension when $A_{ij} = 0$ for $|i-j|>1$~\cite{Cazalilla_Citro_review_11}. The mapping is carried out first using a Holstein-Primakoff transformation~\cite{Holstein_Primakoff_40}, followed by a Jordan-Wigner transformation~\cite{Jordan_Wigner_28},
\begin{equation}\label{eq:mapping}
\hat b_j^\dagger=\hat f^\dagger_j\prod_{m=1}^{j-1}e^{-i\pi \hat f_m^\dagger \hat f^{}_m}\,, \quad
\hat b^{}_j=\prod_{m=1}^{j-1}e^{i\pi \hat f_m^\dagger \hat f^{}_m}\hat f^{}_j\,.
\end{equation}
Using properties of Slater determinants, one can calculate (in polynomial time) the matrix elements of the one-body operators $\hat b^{\dagger}_i \hat b^{}_j$ in the many-body eigenstates $\{|\alpha^{\rm HCB}\rangle\}$ of Eq.~\eqref{eq:Hhcb}, $\langle\alpha^{\rm HCB}|\hat b^{\dagger}_i \hat b^{}_j|\beta^{\rm HCB}\rangle$~\cite{Rigol_Muramatsu_04, Rigol_Muramatsu_05}. This allows one to also compute the matrix elements of the occupation of quasimomentum modes
\begin{equation}\label{eq:mkhcb}
    \hat m_k = \frac{1}{L} \sum_{j,l=1}^L e^{ik(j-l)} \hat b_j^\dagger \hat b^{}_l \;.
\end{equation}
We note that, in order to avoid confusion, we denote the hard-core boson occupation of quasimomentum modes as $\hat m_k$, and the noninteracting spinless fermion occupation of quasimomentum modes as $\hat{\mathsf{m}}_k$.

Because of the hard-core interactions, which are encoded in the nonlocal nature of the mapping between hard-core bosons and noninteracting fermions, the one-body sector of the former system is fundamentally different from the one of the latter, see, e.g., Ref.~\cite{wright_rigol_14} for a comparison of their dynamics. In particular, the occupation of quasimomentum modes is in general different for hard-core bosons and noninteracting fermions, both in equilibrium and out of equilibrium~\cite{Cazalilla_Citro_review_11}. 

More importantly for the purpose of this study, the off-diagonal matrix elements $\langle\alpha^{\rm HCB}|\hat b^{\dagger}_i \hat b^{}_j|\beta^{\rm HCB}\rangle$ need not be sparse as they are for noninteracting fermions. To show it, let us rewrite $\hat m_k$ in Eq.~\eqref{eq:mkhcb} in terms of spinless fermions operators
\begin{equation}\label{eq:mkhcbfer}
    \hat m_k = \frac{1}{L} \sum_{j,l=1}^L e^{ik(j-l)} \hat f^\dagger_j \left(\prod_{m=j}^{l-1}e^{i\pi \hat f_m^\dagger \hat f^{}_m}\right) \hat f^{}_{l} \,.
\end{equation}
Equation~\eqref{eq:mkhcbfer} shows that $\hat m_k$ is a many-body operator in the spinless fermion representation. As a result, it can connect exponentially many many-body eigenstates of the noninteracting spinless fermion Hamiltonian to which the hard-core bosons are mapped.

Our goal is to gain an accurate understanding of the properties of matrix elements of few-body observables in energy eigenstates of integrable interacting models via the computational study of the properties of the matrix elements of $\hat m_k$. The latter can be done efficiently using the mapping onto noninteracting fermions.

\section{Translationally invariant hard-core bosons} \label{sec:HCBstranslation}

We first consider the case in which the hard-core boson Hamiltonian is translationally invariant (no inhomogeneity and periodic boundary conditions):
\begin{equation}
    \label{eq:Hhcb_pw}
    \hat H^{\rm HCB}_{\rm TI}=-\sum_{i=1}^{L-1}( \hat b^\dagger_i\hat b_{i+1}+{\rm H.c.})-(\hat b^\dagger_1\hat b_{L}+{\rm H.c.})\,,
\end{equation}
for which the corresponding spinless-fermion Hamiltonian after the mapping in Eq.~(\ref{eq:mapping}) is
\begin{equation}\label{eq:Hsf_pw}
\hat H^{\rm SF}_{\rm TI}=-\sum_{i=1}^{L-1}( \hat f^\dagger_i\hat f_{i+1}+{\rm H.c.})+(-1)^{N}(\hat f^\dagger_1\hat f_{L}+{\rm H.c.})\,.
\end{equation}
In the latter model, periodic (anti-periodic) boundary conditions are needed for an odd (even) number $N$ of particles. We study systems at quarter filling $N=L/4$ in this section, and consider energy eigenstates with total quasimomentum $\kappa = \sum_{\alpha=1}^{N}\kappa_\alpha=2\pi/L$, where $\kappa_\alpha$ is the quasimomentum of the single-particle eigenstates that are part of the Slater determinant of the many-body eigenstates. We note that, as $L\rightarrow\infty$, $\kappa\rightarrow 0$. We focus on this sector, as opposed to the one with $\kappa=0$, to avoid the parity symmetry present in the latter. We also note that we do not study the half-filled case as it has an additional particle-hole symmetry. For the system sizes considered here, the Hilbert space dimension of the quasimomentum sectors is $D\simeq{L \choose{N}}/L$. This is the Hilbert space dimension that we use in our calculations.

\begin{figure}[!t]
\begin{center}
\includegraphics[width=\columnwidth]{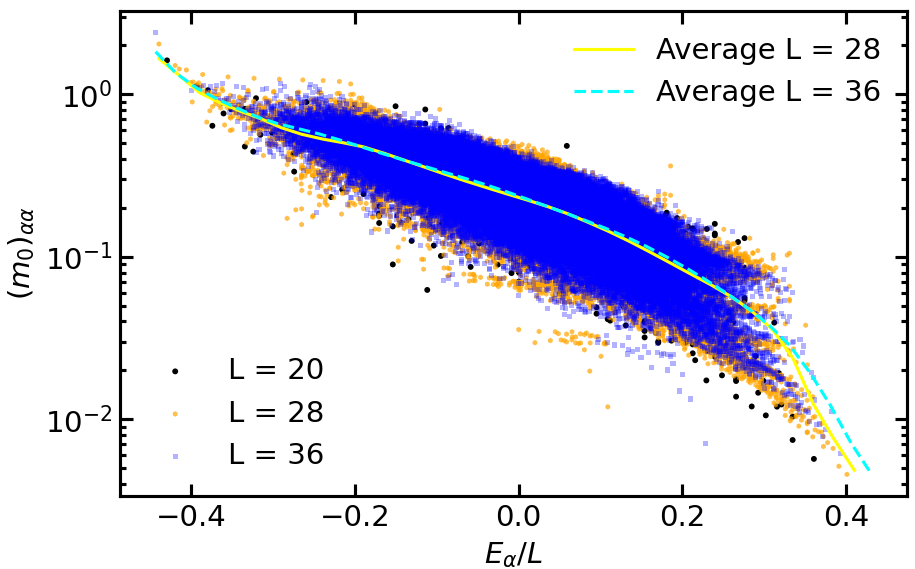}
\caption{\label{fig:PWDiagonal}Diagonal matrix elements $(m_0)_{\alpha\alpha}$ in the energy eigenstates of translationally invariant hard-core bosons in the sector with total quasimomentum $\kappa=2\pi/L$. We show results for systems with $L=20$ (black circles, all matrix elements), $L=28$ (orange hexagons, all matrix elements), and $L=36$ (blue squares; only 1 of every 25 matrix elements) at quarter filling $N=L/4$. The solid (dashed) line shows the average of $(m_0)_{\alpha\alpha}$ within energy windows with $\Delta E_\alpha/L=0.05$ for $L=28$ ($L=36$).} 
\end{center}
\end{figure}

In Fig.~\ref{fig:PWDiagonal}, we show the diagonal matrix elements $(m_0)_{\alpha\alpha}$ of the zero quasimomentum occupation operator $\hat m_0\equiv \hat m_{k=0}$ in Eq.~(\ref{eq:mkhcb}) as a function of the eigenenergy density $E_\alpha/L$ for three different system sizes. They exhibit a well known property of the diagonal matrix elements of integrable models~\cite{rigol_dunjko_08, cassidy_clark_11, vidmar16, Mierzejewski_Vidmar_20}, namely, the support of the matrix elements at any given value of $E_\alpha/L$ does not shrink with increasing the system size. The solid and dashed lines in Fig.~\ref{fig:PWDiagonal} show results for the averages over energy windows with $\Delta E_\alpha/L=0.05$ in the two largest system sizes. The averages overlap (are well converged) for those systems sizes, for which we are able to compute all the matrix elements. 

Even though the support of the matrix elements does not decrease with increasing system size, the variance does decrease~\cite{Biroli_Kollath_10, Ikeda_Watanabe_13, Alba_15}. In order to study the scaling of the variance with increasing system size, and the distribution of the diagonal matrix elements, we carry out calculations in much larger system sizes than the ones shown in Fig.~\ref{fig:PWDiagonal}. For those systems sizes ($L>36$), we cannot compute all the matrix elements so we sample them. 

\begin{figure}[!t]
\begin{center}
\includegraphics[width=0.98\columnwidth]{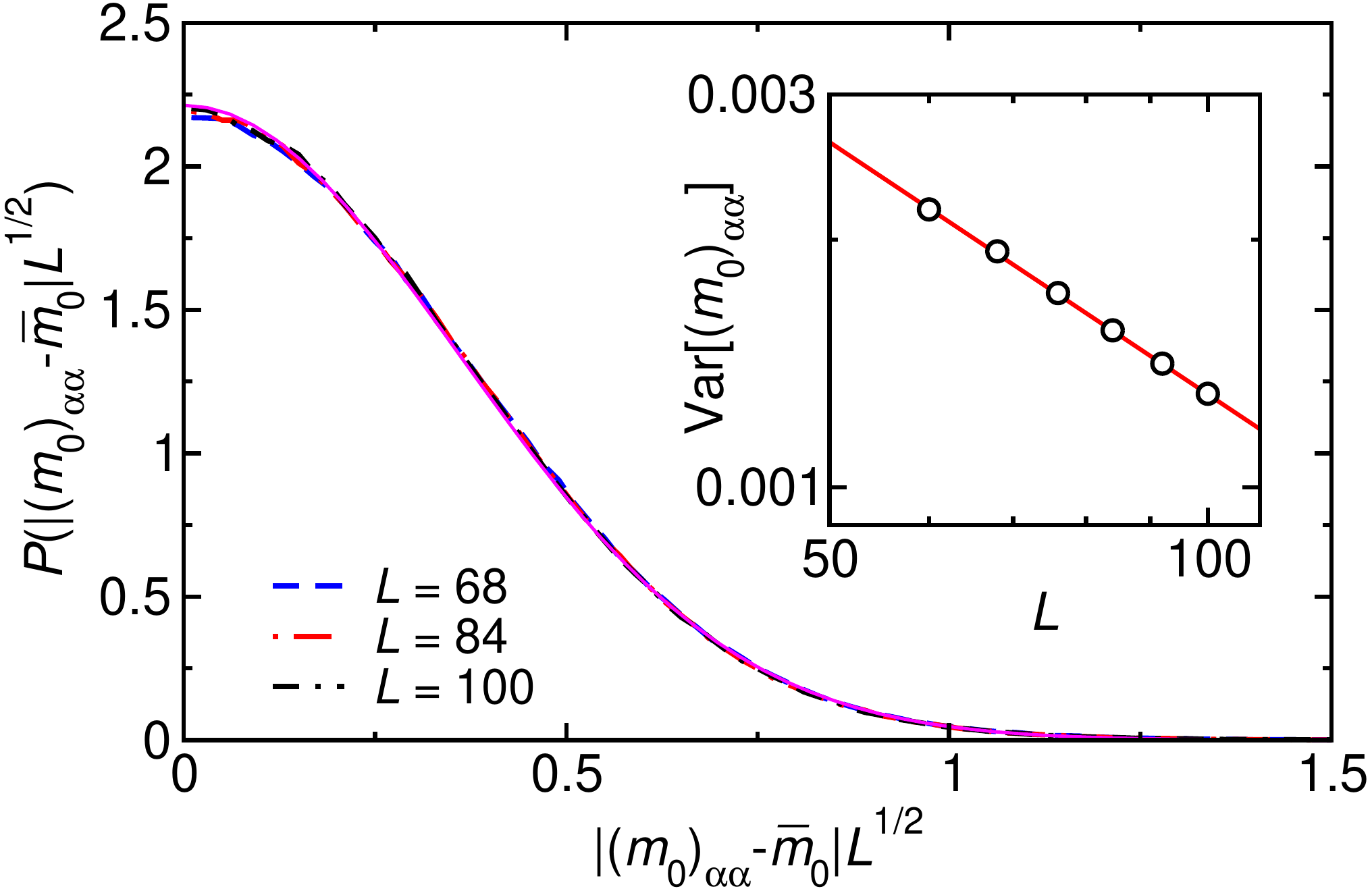}
\caption{\label{fig:DiagPW}Probability density function $P$ of the scaled diagonal matrix elements $|(m_0)_{\alpha\alpha} - \overline{(m_0)}_\alpha|L^{1/2}$, for $L=68$ (dashed line), $L=84$ (dashed-dotted line), and $L=100$ (double dashed-dotted line). In axes labels we simplify $\overline{(m_0)}_\alpha \to  \overline{m}_0$. The solid line is a Gaussian probability density function $P(x) = \frac{2}{\sigma} \sqrt{\frac{1}{2\pi}} e^{-x^2/2\sigma^2}$, where $\sigma=0.36$ is the square root of the variance obtained for $L=100$ [see Eq.~\eqref{def_variance_diag}]. (Inset) The variance [see Eq.~\eqref{def_variance_diag}] plotted as a function of the system size. The solid line is a power-law fit $\propto L^{-\alpha_0}$, where $\alpha_0=1.02$. The numerical results were obtained using $10^6$ eigenstates randomly sampled with $|E_\alpha|/L\leq 10^{-4}$. The moving average $\overline{(m_0)}_\alpha$ is computed averaging $(m_0)_{\alpha\alpha}$ over the 2000 states obtained in the sampling process whose energy is closest to $E_\alpha$ (see text).} 
\end{center}
\end{figure}

In the inset of Fig.~\ref{fig:DiagPW}, we show the variance 
\begin{equation} \label{def_variance_diag}
    {\rm Var}[(m_0)_{\alpha\alpha}] = \frac{1}{|{\cal M}|} \sum_{\alpha\in {\cal M}} [(m_0)_{\alpha\alpha}- \overline{(m_0)}_\alpha]^2 \;,
\end{equation}
where the sum is computed over a set ${\cal M}$ of states at the center of the energy spectrum sampled within an energy window in which $|E_\alpha|/L\leq10^{-4}$. We stress that the variance in Eq.~\eqref{def_variance_diag} is computed with respect to a moving average $\overline{(m_0)}_\alpha$, not with respect to the average in the entire energy window. This is done in order to remove the structure of $(m_0)_{\alpha\alpha}$ as a function of the energy~\cite{lydzba_zhang_21}. Our moving averages $\overline{(m_0)}_\alpha$ are computed over the 2000 states obtained in the sampling process whose energy is closest to $E_\alpha$. A power-law fit to those numerical results shows that the variance decreases $\propto L^{-1}$, i.e., it vanishes in the thermodynamic limit~\cite{Biroli_Kollath_10, Ikeda_Watanabe_13, Alba_15}. This is to be contrasted with the much faster scaling in nonintegrable (quantum-chaotic interacting) systems, in which the variance vanishes exponentially fast in the system size (see, e.g., Ref.~\cite{LeBlond_Mallayya_19} for a recent comparison between numerical results obtained in integrable and nonintegrable spin-1/2 XXZ chains).

In Fig.~\ref{fig:DiagPW} we show the probability density function (PDF) of the scaled matrix elements $|(m_0)_{\alpha\alpha}- \overline{(m_0)}_\alpha| L^{1/2}$. We define the PDF, $P$, of a variable $x$ in an interval $[x,x+\Delta x]$ as
\begin{equation}
    P(x) = \frac{1}{\cal N} \frac{\Delta \cal N}{\Delta x} \;,
\end{equation}
where $\cal N$ is the total number of elements ($\Delta \cal N$ is the number of elements in $[x,x+\Delta x]$). Figure~\ref{fig:DiagPW} shows that $P(|(m_0)_{\alpha\alpha}- \overline{(m_0)}_\alpha| L^{1/2})$ is a system-size-independent Gaussian. The same Gaussian behavior was found in Ref.~\cite{Alba_15} for the diagonal matrix elements of elements of reduced density matrices in eigenstates of the integrable spin-1/2 isotropic Heisenberg chain.

Having shown that the properties of the diagonal matrix elements of $\hat m_0$ are qualitatively similar to those observed in integrable interacting systems that are not mappable onto noninteracting models, we turn our attention to the properties of the off-diagonal matrix elements $(m_0)_{\alpha\beta}$. As for the diagonal matrix elements, we consider only matrix elements between eigenstates within the total quasimomentum sector $\kappa = 2\pi/L$. 

The variance of the off-diagonal matrix elements (whose average is negligibly small) is
\begin{equation} \label{def_variance}
    {\rm Var}[(m_0)_{\alpha\beta}] = \frac{1}{|\cal M'|} \sum_{\alpha,\beta \in {\cal M'}} |(m_0)_{\alpha\beta}|^2 \;.
\end{equation}
We carry out our calculations over a set ${\cal M'}$ of pairs of eigenstates $|\alpha\rangle,\, |\beta\rangle$ with $\bar E_{\alpha\beta} = (E_\alpha+E_\beta)/2$ at the center of the energy spectrum, namely, in a small window of energy $|\bar E_{\alpha\beta} - \bar E_0| \leq \Delta E/2$, where $\bar E_0 = {\rm Tr}\{\hat H\}/D$ ($E_0 = 0$ in the translationally invariant model considered in this section). In addition to their average energy $\bar E_{\alpha\beta}$, pairs of eigenstates can be labeled by their energy difference $\omega_{\alpha\beta} = E_\alpha-E_\beta$.  We coarse grain ${\rm Var}[(m_0)_{\alpha\beta}]$ so that $|\omega_{\alpha\beta} - \omega| \leq \Delta \omega/2$. We quote the specific widths $\Delta E$ and $\Delta \omega$ used in the calculations in the caption of each figure. Finally, we report in our plots the scaled variance
\begin{equation}\label{eq:scaledvar}
    V_{m_0}(0, \omega)=D \,{\rm Var}[(m_0)_{\alpha\beta}] \,,
\end{equation}
which, given the fact that observables have a fixed (properly normalized) Hilbert-Schmidt norm, is the quantity that is expected to remain finite in the thermodynamic limit ($D \to \infty$).

\begin{figure}[!t]
\begin{center}
\includegraphics[width=0.98\columnwidth]{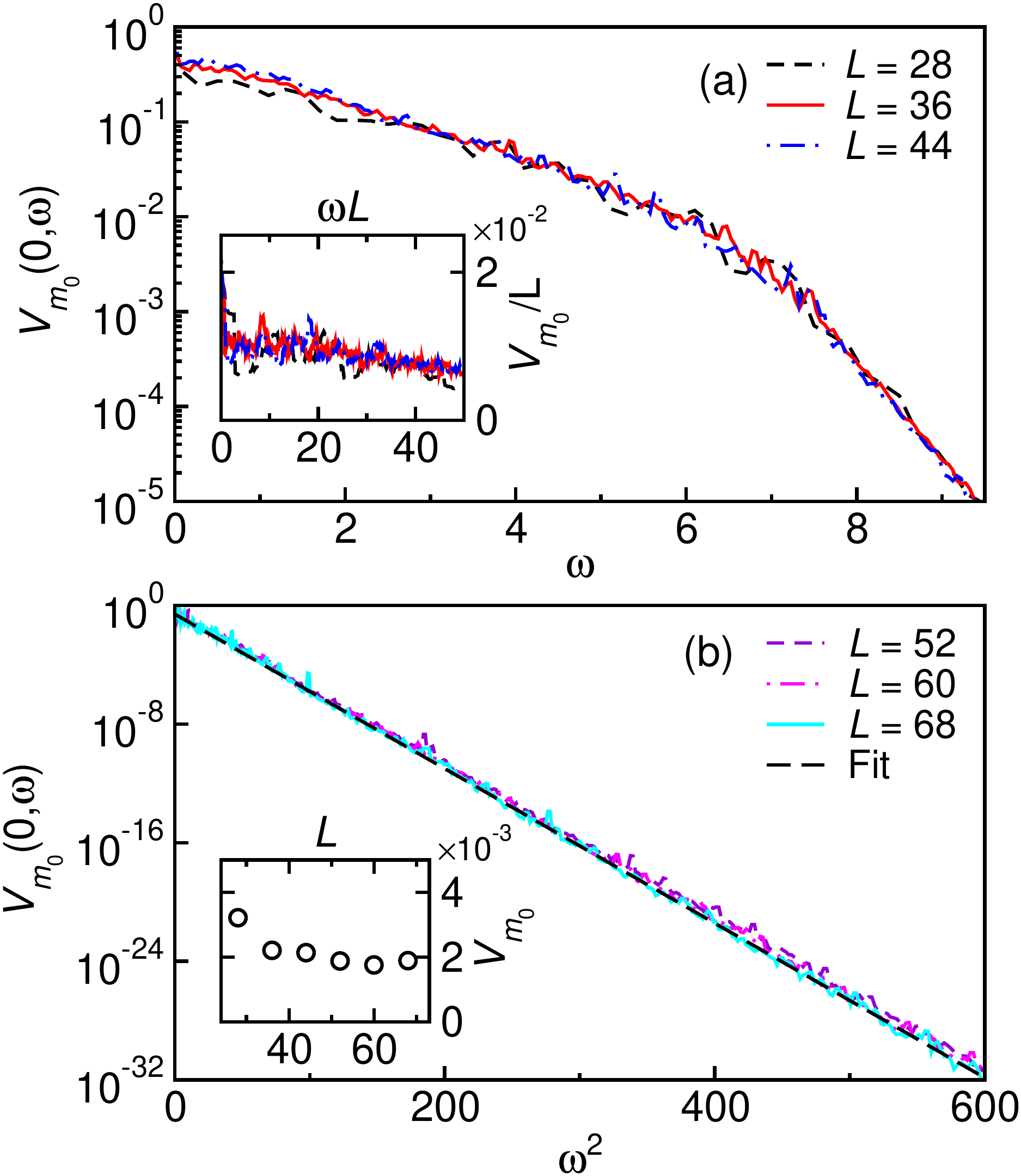}
\caption{\label{fig:PWvariance}Scaled variance $V_{m_0}(0,\omega)$ [Eq.~(\ref{eq:scaledvar}] of the off-diagonal matrix elements $(m_0)_{\alpha\beta}$ in the translationally invariant hard-core boson model at the center of the energy spectrum. The energy eigenstates are from the $\kappa=2\pi/L$ total quasimomentum sector of systems at quarter filling ($N=L/4$). (a) $V_{m_0}(0,\omega)$ plotted as a function of $\omega$ at low and intermediate frequencies ($\omega \in [0,9]$). We show results for systems with sizes $L=28$ (dashed line), 36 (solid line), and 44 (dashed-dotted line). (Inset) Rescaled $V_{m_0}(0,\omega)/L$ plotted as a function of $\omega L$ at low frequencies. (b) $V_{m_0}(0,\omega)$ plotted as a function of $\omega^2$ ($\omega \lesssim 25$) for systems with sizes $L=52$ (dashed line), 60 (dashed-dotted line), and 68 (solid line). The straight dashed line is a Gaussian fit $\propto e^{-a \omega^2}$ to the $L=68$ results for $\omega^2\in[300,600]$, with the fitting parameter $a=0.12$. (Inset) $V_{m_0}(0,\omega=7)$ vs the system size. For all the results shown in this figure, $\Delta E/L=2\times10^{-4}$. We compute all pairs of eigenstates in this interval for $L\leq 36$, while for $L\geq 44$ we randomly select at least $6\times10^7$ pairs (see Appendix~\ref{app:sampling}). The variances in the main panels are coarse grained using a $\Delta\omega=0.05$ (except for $L=28$ for which $\Delta\omega=0.2$). We use a finer coarse graining in the inset in (a), $\Delta\omega=0.02$ (except for $L=28$ for which $\Delta\omega=0.1$), and the results are plotted as a running average.}
\end{center}
\end{figure}

The main panel of Fig.~\ref{fig:PWvariance}(a) shows the scaled variance $V_{m_0}(0,\omega)$ as a function of $\omega$ at small and intermediate frequencies. The results for three different system sizes collapse at intermediate frequencies, thereby justifying the use of the scaled variance in Eq.~(\ref{eq:scaledvar}) as a meaningful quantity in the thermodynamic limit. The scaled variance in a wider frequency interval ($\omega \lesssim 25$), for three system sizes larger than those in Fig.~\ref{fig:PWvariance}(a), is shown in Fig.~\ref{fig:PWvariance}(b) as a function of $\omega^2$. The results for the three system sizes collapse at high frequencies, and they are consistent with the Gaussian functional form 
\begin{equation} \label{eq:gaussian}
    V_{m_0}(0,\omega)=A e^{-a\omega^2} \;,    
\end{equation}
where $A$ and $a$ are constants. The variance of the off-diagonal matrix elements of observables at high frequency was also found to exhibit a Gaussian decay in the integrable spin-1/2 XXZ chain~\cite{LeBlond_Mallayya_19}, and in quantum-chaotic interacting models in which integrability is broken by perturbations that are not extensive in the system size~\cite{jansen_stolpp_19, schoenle_jansen_21}.

The behavior of the scaled variance $V_{m_0}(0,\omega)$ is qualitatively different at low frequencies $\omega \propto 1/L$. The inset in Fig.~\ref{fig:PWvariance}(a) shows that, in this regime, the results for the variance collapse only when plotting $V_{m_0}(0,\omega)/L$ vs $\omega L$. Similar behaviors have been observed in some quantum-chaotic interacting models~\cite{dalessio_kafri_16, Brenes_LeBlond_20, brenes_goold_20} and in the integrable spin-1/2 XXZ chain~\cite{LeBlond_Rigol_Eigenstate_20}, and can be attributed to the presence of ballistic transport.

In what follows we the study the distributions of the off-diagonal matrix elements $(m_0)_{\alpha\beta}$ at a fixed frequency $\omega=7$. This frequency is in the intermediate frequency regime in Fig.~\ref{fig:PWvariance}(a), and is sufficiently high so that the matrix elements are not affected by the low-frequency ``ballistic'' scaling seen in the inset in Fig.~\ref{fig:PWvariance}(a). (We report results for the distribution of $(m_0)_{\alpha\beta}$ at low-frequencies in Sec.~\ref{sec:xxz}.) The inset in Fig.~\ref{fig:PWvariance}(b) shows that the variance at $\omega=7$ is, up to small fluctuations, independent of the system size. In Appendix~\ref{app:mkpi}, we show that the occupation of other quasimomentum modes (specifically, of $k=\pi/2$ and $\pi$) exhibit the same qualitative behavior as the one discussed here for $k=0$.

We study the PDFs of the squared absolute value of the scaled matrix elements (which enter in response functions, and others~\cite{dalessio_kafri_16, Mallayya_Rigol_19})
\begin{equation} \label{def_tilde_m0}
    |(\tilde{m}_0)_{\alpha\beta}|^2=|(m_0)_{\alpha\beta}\sqrt{D}|^2 \;.
\end{equation}
To be able to study large systems (with up to $L=100$) so that we can unveil the scaling of the PDFs with the system size, we randomly sample the matrix elements in the targeted $\omega$ window (see Appendix~\ref{app:sampling}). 

\begin{figure}[!t]
\begin{center}
\includegraphics[width=0.99\columnwidth]{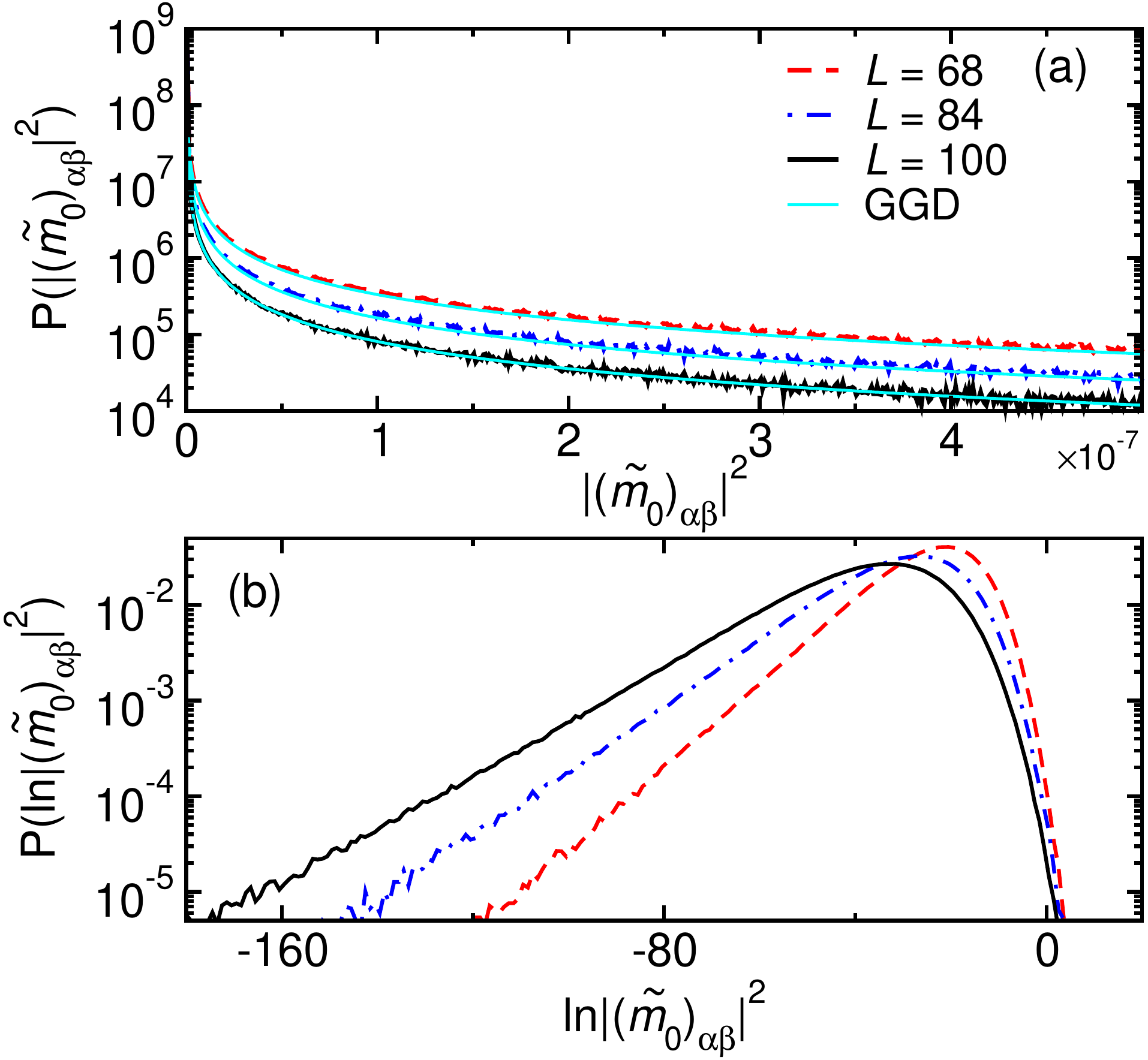}
\caption{\label{fig:PWdistribution}(a) Probability density function $P$ of $|(\tilde{m}_0)_{\alpha\beta}|^2$ [see Eq.~(\ref{def_tilde_m0})] in the translationally invariant hard-core boson model. The thin (cyan) lines overlapping with the results show the prediction of the generalized Gamma distribution (GGD) in Eq.~\eqref{eq:prob1}, with the fitting parameters from Fig.~\ref{fig:PWscale}(b). (b) The same results as in panel (a), but plotted as the probability density function of $\ln{|(\tilde{m}_0)_{\alpha\beta}|^2}$. We study eigenstates in the quasimomentum sector $\kappa = 2\pi/L$ for systems at quarter filling $N = L/4$. We show results for systems with sizes $L=68$ (dashed line), 84 (dashed-dotted line), and 100 (solid line). We randomly select at least $5\times 10^6$ pairs of eigenstates with $\Delta E/L= 2\times 10^{-4}$ and $\Delta\omega=0.05$ about $\omega=7$.} 
\end{center}
\end{figure}

The PDFs $P(|(\tilde{m}_0)_{\alpha\beta}|^2)$ for $L=68,\, 84\,$ and 100 are shown in Fig.~\ref{fig:PWdistribution}(a). They exhibit sharp peaks as $|(\tilde{m}_0)_{\alpha\beta}|^2\rightarrow0$, and long tails for large matrix elements, as those found for local observables in the integrable spin-1/2 XXZ chain~\cite{LeBlond_Mallayya_19, brenes_goold_20, LeBlond_Rigol_Eigenstate_20}. In Fig.~\ref{fig:PWdistribution}(b), we plot the corresponding PDFs $P(\ln |(\tilde{m}_0)_{\alpha\beta}|^2)$. They exhibit the skewed log-normal like shape observed in Refs.~\cite{LeBlond_Mallayya_19, brenes_goold_20, LeBlond_Rigol_Eigenstate_20}, and clearly visible tails for small matrix elements that were visible only in some instances in the much smaller system sizes studied in Refs.~\cite{LeBlond_Mallayya_19, brenes_goold_20, LeBlond_Rigol_Eigenstate_20}. Both plots make apparent that those distributions are not independent of the system size (as they would be for a Gaussian, for which the mean and the variance fix all the higher moments). In particular, the peak in $P(|(\tilde{m}_0)_{\alpha\beta}|^2)$ as $|(\tilde{m}_0)_{\alpha\beta}|^2\rightarrow0$ sharpens, while $P(\ln |(\tilde{m}_0)_{\alpha\beta}|^2)$ exhibits a maximum that drifts to lower values of $|(\tilde{m}_0)_{\alpha\beta}|^2$ with increasing the system size.

\begin{figure}[!t]
\begin{center}
\includegraphics[width=0.99\columnwidth]{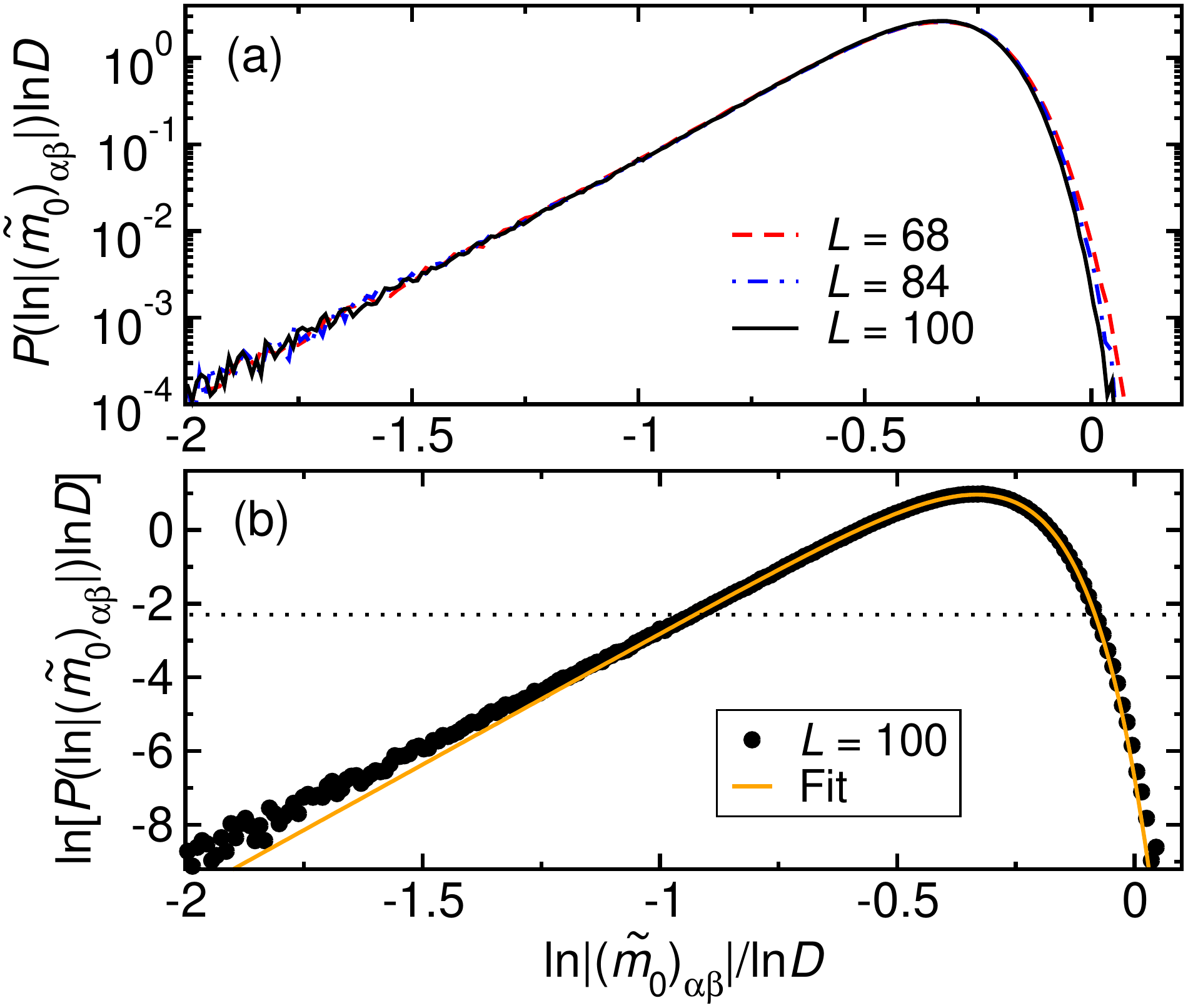}
\caption{\label{fig:PWscale}(a) Rescaled probability density function $P(\ln|(\tilde{m}_0)_{\alpha\beta}|) \ln D$ as a function of $\ln |(\tilde{m}_0)_{\alpha\beta}|/\ln D$ in the translationally invariant hard-core boson model. The numerical results are the same as in Fig.~\ref{fig:PWdistribution}. (b) The symbols show the logarithm of the results for $L=100$ in (a), and the solid line is a fit to the points above the dotted line [$P(\ln|(\tilde{m}_0)_{\alpha\beta}|) \ln D \geq 0.1$] using the function in Eq.~(\ref{eq:fit1}). The fitting parameters are $A_0=4.30$, $B_0=7.29$, $k_0=7.11$, and $x_0=-0.33$.} 
\end{center}
\end{figure}

These results suggest that further rescaling of the matrix elements as a function of $D$ is needed if one is to find a PDF that is meaningful in the thermodynamic limit. We rescale
\begin{equation}\label{eq:rescm0}
    \ln |(\tilde{m}_0)_{\alpha\beta}|^2  \to \frac{\ln |(\tilde{m}_0)_{\alpha\beta}|^2}{\ln D^2}= \frac{\ln |(\tilde{m}_0)_{\alpha\beta}|}{\ln D}\;,
\end{equation}
and, consequently (to ensure the new distribution is normalized),
\begin{eqnarray}
    P(\ln |(\tilde{m}_0)_{\alpha\beta}|^2)  &\to& P(\ln |(\tilde{m}_0)_{\alpha\beta}|^2) \; \ln D^2\nonumber\\&&= P(\ln |(\tilde{m}_0)_{\alpha\beta}|) \; \ln D \;. \label{def_Pln_rescaling}
\end{eqnarray}
Figure~\ref{fig:PWscale}(a) shows that this yields a very good collapse of the results for different values of $L$, specially about and below the maximum of $P(\ln |(\tilde{m}_0)_{\alpha\beta}|)$. The collapse degrades at the highest values of $|(\tilde{m}_0)_{\alpha\beta}|$, for which $P(\ln |(\tilde{m}_0)_{\alpha\beta}|)$ exhibits a sharp decrease. Properly sampling that part of the distribution becomes increasingly challenging with increasing system size.

The behavior in Fig.~\ref{fig:PWscale}(a) is consistent with the logarithm of the PDF [plotted in Fig.~\ref{fig:PWscale}(b) for $L=100$] being linear for small values of $\ln |(\tilde{m}_0)_{\alpha\beta}|/\ln D$, and exponential for large values of $\ln| (\tilde{m}_0)_{\alpha\beta}|/\ln D$. We therefore fit the results in Fig.~\ref{fig:PWscale}(b) to the function
\begin{align}\label{eq:fit1}
\ln[P(&\ln{|(\tilde{m}_0)_{\alpha\beta}}|)\ln{D}]=\\\nonumber&A_0+k_0 \frac{\ln|(\tilde{m}_0)_{\alpha\beta}|}{\ln D} -\exp{\left[B_0\left( \frac{\ln|(\tilde{m}_0)_{\alpha\beta}|}{\ln D}-x_0\right)\right]}\,,
\end{align}
with $A_0$, $B_0$, $k_0$, and $x_0$ being fitting parameters. The fit provides an excellent description of the data in the regime in which the results for different system sizes exhibit a collapse in Fig.~\ref{fig:PWscale}(a).

The corresponding distribution for $|(\tilde{m}_0)_{\alpha\beta}|^2$ is
\begin{equation}\label{eq:prob1}
    P(|(\tilde{m}_0)_{\alpha\beta}|^2)=P_D |(\tilde{m}_0)_{\alpha\beta}|^{2(k_D-1)}\exp\left[-\alpha_D |(\tilde{m}_0)_{\alpha\beta}|^{2B_D}\right]\,,
\end{equation} 
with $P_D=\exp[A_0]/(2\ln D)$, $k_D=k_0/(2\ln D)$, $\alpha_D=\exp(-B_0x_0)$, and $B_D=B_0/(2\ln D)$. The distribution in Eq.~(\ref{eq:prob1}) in known as the generalized Gamma distribution~\cite{gamma_distribution}. In Fig.~\ref{fig:PWdistribution}(a), we show that it fits well the results for $P(|(\tilde{m}_0)_{\alpha\beta}|^2)$ for different system sizes.

The results in this section open two important questions that we address in the reminder of this paper. The first one is whether perturbing the hard-core boson model considered, e.g., by breaking translational invariance, still results in PDFs of the off-diagonal matrix elements that are described by generalized Gamma distributions. If yes, we need to understand whether the parameters of the distributions depend on Hamiltonian parameters. The second question is what happens if the hard-core boson model undergoes a localization transition. In order to address these questions, we consider next the Aubry-Andr\'e model.

\section{Aubry-Andr\'e model for\\ spinless fermions} \label{sec:fermionsAA}

The Aubry-Andr\'e model is a paradigmatic model of a delocalization-localization transition in one-dimensional lattices~\cite{aubry1980analyticity}. For open boundary conditions, the Aubry-Andr\'e model Hamiltonian for noninteracting spinless fermions can be written as
\begin{equation}\label{eq:HsfAA}
\hat H^{\rm SF}_{\rm AA}=-J\sum_{i=1}^{L-1}( \hat f^\dagger_i\hat f^{}_{i+1}+{\rm H.c.})+\lambda J \sum_{i=1}^L \cos(2\pi\beta i+\phi_0) \hat f^\dagger_i\hat f^{}_i\,,
\end{equation}
where $J$ is the hopping energy between nearest neighbor sites, and the on-site potential has a quasiperiodic functional form with a magnitude $\lambda J$, incommensurate period $1/\beta$ (we choose $\beta$ to be the inverse golden mean $\beta=(\sqrt{5}-1)/2$, considered to be the most irrational number~\cite{Sokoloff_85}), and a global phase shift $\phi_0$. We set $J=1$ in what follows. The single-particle eigenstates of the Aubry-Andr\'e model have a delocalization-localization transition at $\lambda_c=2$~\cite{aubry1980analyticity}. For $\lambda<\lambda_c$, all single-particle eigenstates are extended, while for $\lambda>\lambda_c$ they are localized. At the transition point $\lambda_c$, the energy spectrum exhibits the well known Hofstadter butterfly fractal structure~\cite{Hofstadter_76}. 

\begin{figure}[!t]
\begin{center}
\includegraphics[width=0.99\columnwidth]{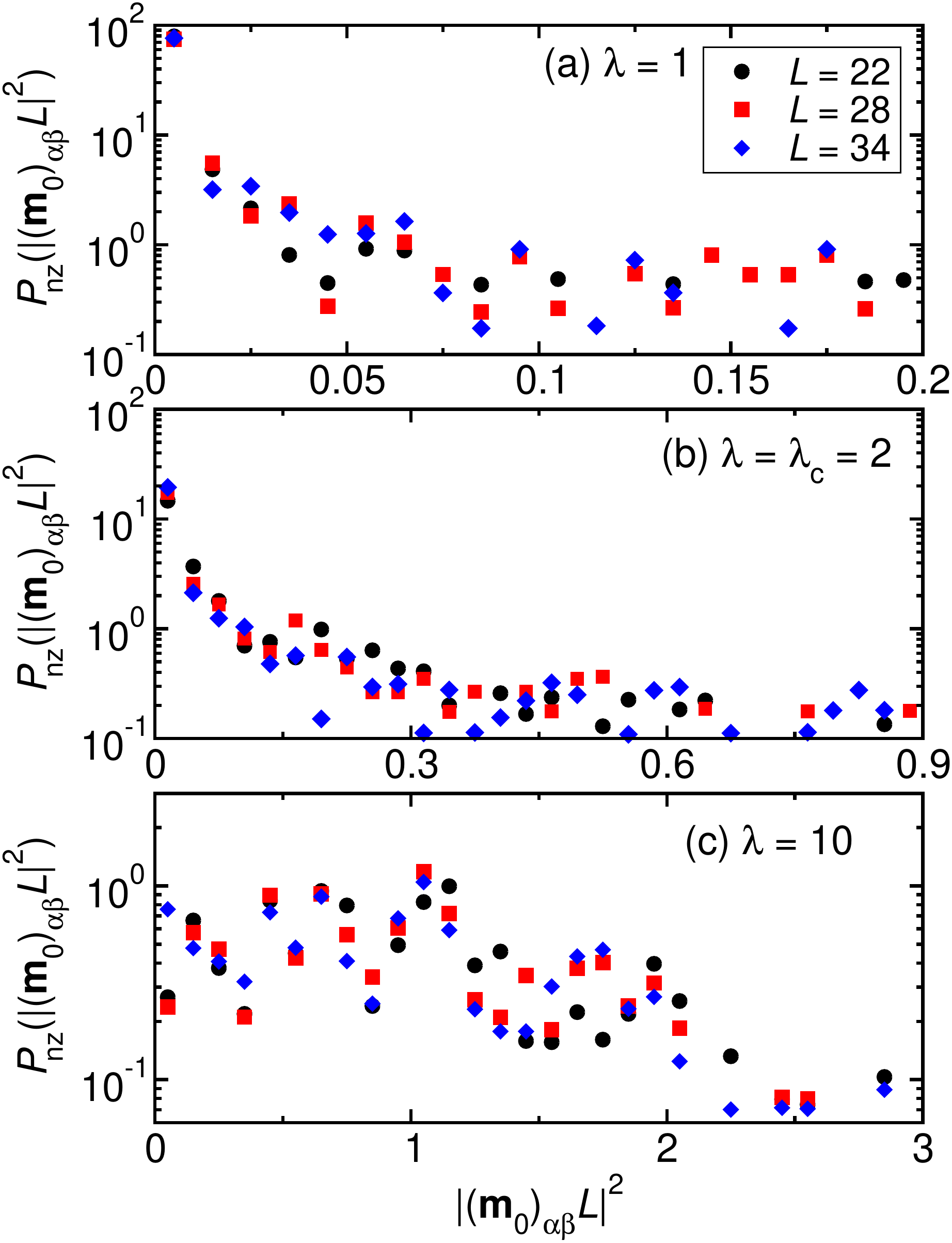}
\caption{\label{fig:Fermiondistribution}Probability density function $P_{\rm nz}$ of the scaled nonzero off-diagonal matrix elements $|(\mathsf{m}_0)_{\alpha\beta}L|^2$ in the spinless fermion Aubry-Andr\'e model, with a phase shift $\phi_0=0$. The systems studied have $L=22$ (black circles), 28 (red squares), and 34 (blue diamonds), and are at half filling ($N=L/2$). Results are shown for (a) $\lambda=1$ (delocalized regime), (b) $\lambda=2$ (transition point), and (c) $\lambda=10$ (localized regime). We consider all pairs of eigenstates with $\Delta E/L= 2\times 10^{-4}$.}
\end{center}
\end{figure}

In contrast to translationally invariant systems in which all the off-diagonal matrix elements of $\hat{\mathsf{m}}_0$ vanish in the many-body eigenstates of the Hamiltonian (because all the single-particle eigenstates are quasimomentum eigenstates), this is not the case in the Aubry-Andr\'e model. As follows from the discussion in Sec.~\ref{sec:fermionsgeneral}, the off-diagonal matrix elements $(\mathsf{m}_0)_{\alpha\beta}$ in the Aubry-Andr\'e model must still be sparse, i.e., the overwhelming majority of them vanish. The magnitude of those that are nonzero is expected to scale with the system size according to Eq.~(\ref{m0_typical}), i.e., $|(\mathsf{m}_0)_{\alpha\beta}|^2 \propto 1/L^2$. Therefore, here we study the PDF of the nonzero matrix elements $P_{\rm nz}$ as a function of scaled matrix elements $|(\mathsf{m}_0)_{\alpha\beta} L|^2$. 

\begin{figure}[!t]
\begin{center}
\includegraphics[width=0.99\columnwidth]{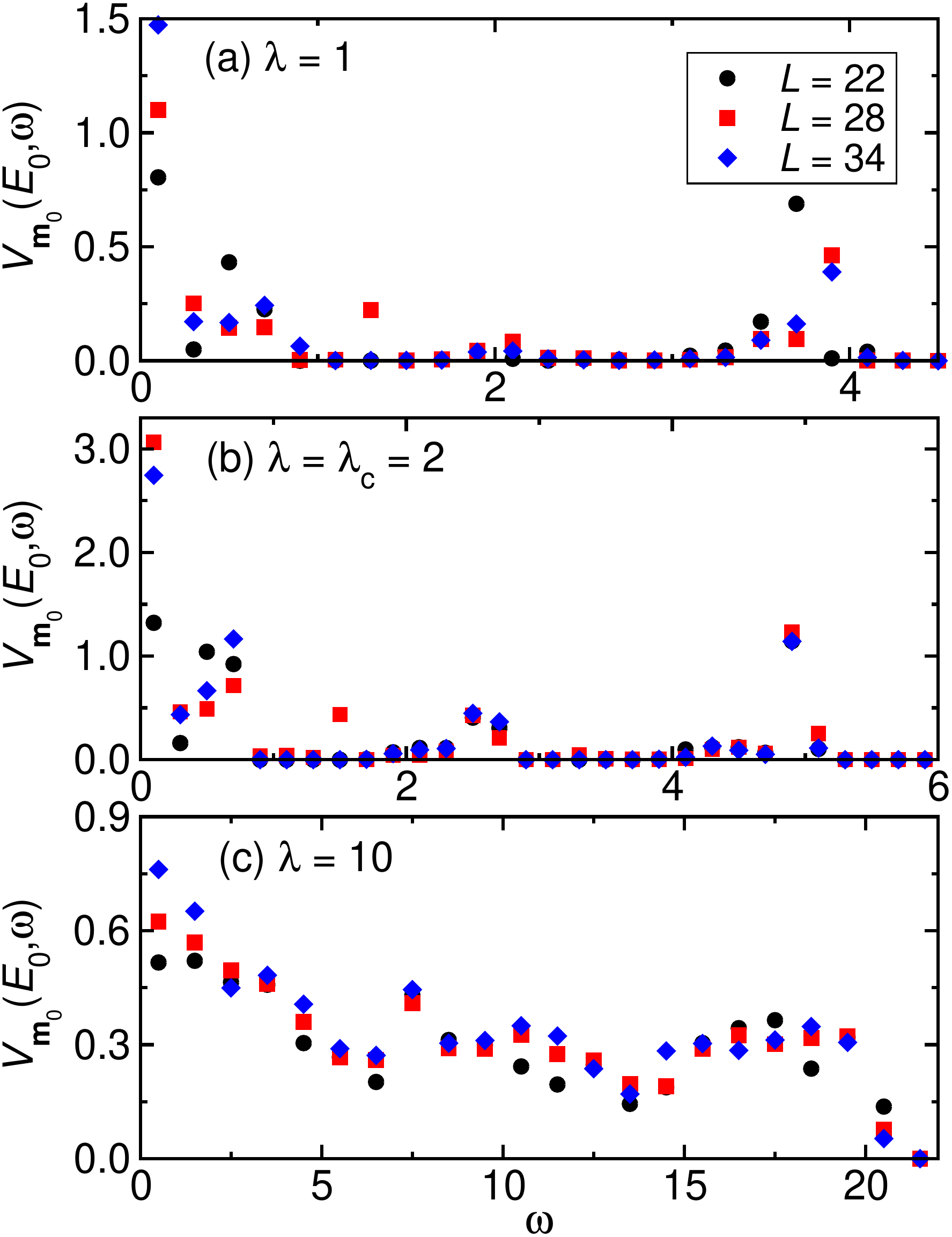}
\caption{\label{fig:Fermionvariance} Scaled variance $V_{\mathsf{m}_0}(E_0,\omega)$ of the off-diagonal matrix elements $(\mathsf{m}_0)_{\alpha\beta}$ in the spinless fermion Aubry-Andr\'e model, with a phase shift $\phi_0=0$. The systems studied have $L=22$ (black circles), 28 (red squares), and 34 (blue diamonds), and are at half filling ($N=L/2$). Results are shown for (a) $\lambda=1$ (delocalized regime), (b) $\lambda=2$ (transition point), and (c) $\lambda=10$ (localized regime). We consider all pairs of eigenstates with $\Delta E/L= 2\times 10^{-4}$, and average the results over frequency windows $\Delta\omega=0.2$ in (a) and (b), and $\Delta\omega=1.0$ in (c).}
\end{center}
\end{figure}

Results for $P_{\rm nz}(|(\mathsf{m}_0)_{\alpha\beta} L|^2)$ are shown in Fig.~\ref{fig:Fermiondistribution}, at $\lambda=1$ in the delocalized regime [Fig.~\ref{fig:Fermiondistribution}(a)], at the transition point $\lambda = \lambda_c$ [Fig.~\ref{fig:Fermiondistribution}(b)], and at $\lambda=10$ in the localized regime [Fig.~\ref{fig:Fermiondistribution}(c)]. In all cases one can see that, up to fluctuations, the scaled distributions collapse for different system sizes. The PDFs exhibit a sharp peak as $(\mathsf{m}_0)_{\alpha\beta} L \rightarrow 0$ for $\lambda\leq\lambda_c$ (all of them vanish for $\lambda=0$), which broadens and becomes a broad distribution upon increasing $\lambda$ for $\lambda>\lambda_c$. 

Next we compute the scaled variance $V_{\mathsf{m}_0}(\bar E_0, \omega)$, defined for the fermions as in Eq.~(\ref{eq:scaledvar}) for the hard-core bosons. This is the quantity that is expected to remain finite in the thermodynamic limit. In Fig.~\ref{fig:Fermionvariance}, we plot $V_{\mathsf{m}_0}(\bar E_0, \omega)$ for the same values of $L$ and $\lambda$ as in Fig.~\ref{fig:Fermiondistribution}. The results for $V_{\mathsf{m}_0}(\bar E_0, \omega)$ in different systems sizes collapse (up to fluctuations), which suggests that $V_{\mathsf{m}_0}(\bar E_0, \omega)$ is a well-defined function in the thermodynamic limit. Its functional form depends strongly on whether $\lambda$ is below or above the localization transition. One can also see in Fig.~\ref{fig:Fermionvariance} that $V_{\mathsf{m}_0}(\bar E_0, \omega)$ as a function of $\omega$ is qualitatively different from $V_{{m}_0}(0, \omega)$ as a function of $\omega$ for hard-core bosons (see Fig.~\ref{fig:PWvariance}). For noninteracting spinless fermions the variance is nonzero (and so are the off-diagonal matrix elements) for an $\omega$ range that is determined by the bandwidth of the single-particle spectrum, and no Gaussian decay occurs for large values of $\omega$.

\section{Aubry-Andr\'e model for\\ hard-core bosons} \label{sec:HCBsAA}

The Aubry-Andr\'e model for hard-core bosons, with open boundary conditions, can be written as
\begin{equation}\label{eq:HhcbAA}
\hat H^{\rm HCB}_{\rm AA}=-\sum_{i=1}^{L-1}( \hat b^\dagger_i\hat b^{}_{i+1}+{\rm H.c.})+\lambda \sum_{i=1}^L \cos(2\pi\beta i+\phi_0) \hat b^\dagger_i\hat b^{}_i\,,
\end{equation}
and can be mapped onto the spinless fermion Aubry-Andr\'e model in Eq.~(\ref{eq:HsfAA}). As in the previous sections, we focus on the matrix elements of the occupation of the zero quasimomentum mode $\hat m_{0}$.

\begin{figure}[!t]
\begin{center}
\includegraphics[width=0.99\columnwidth]{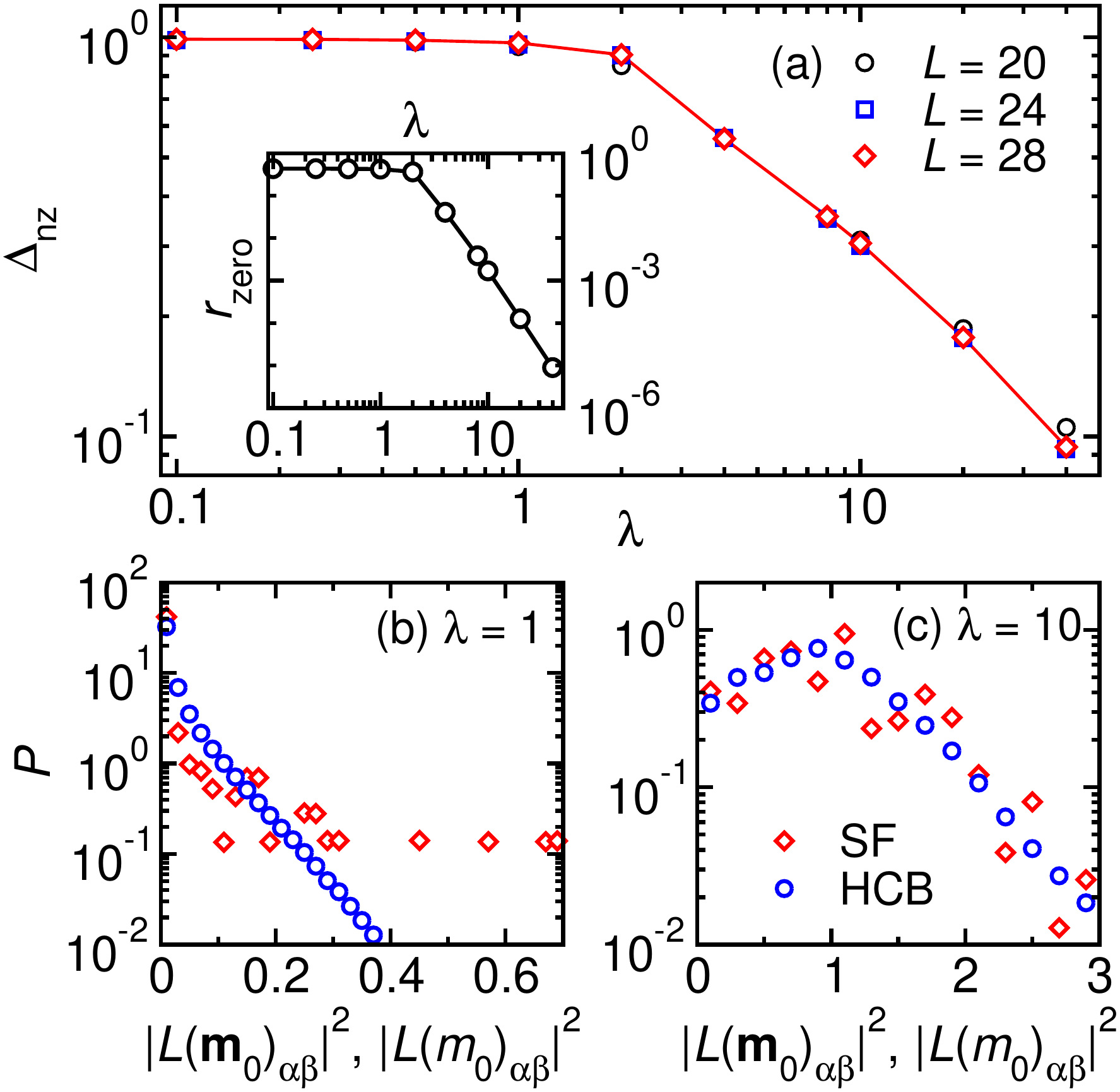}
\caption{\label{fig:Bosondistribution} Comparison between the matrix elements of hard-core bosons and spinless fermions in the Aubry-Andre model at $\phi_0=0$. (a) Relative difference between the off-diagonal matrix elements $\Delta_{\rm nz}$ [see Eq.~(\ref{def_delta_nz})] vs $\lambda$ for three different system sizes. (b, c) PDFs of the scaled matrix elements of spinless fermions $|L({\bf m}_0)_{\alpha\beta}|^2$ (diamonds) and of hard-core bosons $|L(m_0)_{\alpha\beta}|^2$ (circles) at $\lambda=1$ and $10$, respectively, for $L=28$. We consider all pairs of eigenstates for which the matrix elements of the spinless fermions are nonzero, with $\Delta E/L= 2\times 10^{-4}$, in systems at half filling $N=L/2$. Inset in (a): $r_{\rm zero}$ [see Eq.~(\ref{def_rzero})] vs $\lambda$ for $L=20$. To compute this quantity we use only off-diagonal matrix elements between pairs of eigenstates for which the corresponding matrix elements of the spinless fermions are zero.} 
\end{center}
\end{figure}

As advanced in Sec.~\ref{sec:general}, we have seen that the main difference between the off-diagonal matrix elements $(m_0)_{\alpha\beta}$ of hard-core bosons in the translationally invariant model and the matrix elements $(\mathsf{m}_0)_{\alpha\beta}$ of spinless fermions in the Aubry-Andr\'e model is that the overwhelming majority of the former are nonzero. We begin our study of the off-diagonal matrix elements $(m_0)_{\alpha\beta}$ of hard-core bosons in the Aubry-Andr\'e model by computing the relative difference between those that are nonzero for spinless fermions (whose number grows polynomially in the system size) and the same matrix elements for hard-core bosons
\begin{equation} \label{def_delta_nz}
    \Delta_{\rm nz} = \frac{\sum_{\alpha,\beta \in {\rm nz}}||(m_0)_{\alpha\beta}|^2-|(\mathsf{m}_0)_{\alpha\beta}|^2|}
    {\sum_{\alpha,\beta \in {\rm nz}}|(m_0)_{\alpha\beta}|^2+\sum_{\alpha,\beta \in {\rm nz}}|(\mathsf{m}_0)_{\alpha\beta}|^2} \;.
\end{equation}
Again, the sum over $\alpha$ and $\beta$ runs over the pairs of eigenstates for which $(\mathsf{m}_0)_{\alpha\beta}$ are nonzero ($\alpha,\beta \in {\rm nz}$).

Results for $\Delta_{\rm nz}$ vs $\lambda$, for pairs of eigenstates whose average energy is at the center of spectrum, are shown in the main panel of Fig.~\ref{fig:Bosondistribution}(a) for different system sizes (for which we compute all pairs of eigenstates in the selected window). $\Delta_{\rm nz}$ can be seen to be approximately one for $\lambda< 2$, a regime in which (as for the translationally invariant case) we expect the off-diagonal matrix elements of the hard-core bosons to be dense, while the off-diagonal matrix elements of the spinless fermions are sparse. Because of the fixed Hilbert-Schmidt norm, their magnitude must scale differently with the system size (for the former it should be negligible when compared to the latter), and that results in $\Delta_{\rm nz} \approx 1$. For $\lambda< 2$ the off-diagonal matrix elements of hard-core bosons and spinless fermions also exhibit very different PDFs. This can be seen in Fig.~\ref{fig:Bosondistribution}(b), where we show the PDFs for $\lambda=1$ for pairs of eigenstates $\alpha$ and $\beta$ for which $(\mathsf{m}_0)_{\alpha\beta}$ are nonzero.

Figure~\ref{fig:Bosondistribution}(a) also shows that for $\lambda> 2$, in the localized regime, $\Delta_{\rm nz} \to 0$ as $\lambda$ increases. Namely, the off-diagonal matrix elements of the hard-core bosons approach the values of the off-diagonal matrix elements of the fermions. As a result, one concludes that the off-diagonal matrix elements of the hard-core bosons become sparse. In this regime localization precludes $\hat m_0$ from connecting an exponentially large number of eigenstates. Figure~\ref{fig:Bosondistribution}(c) shows that in this regime, specifically for $\lambda=10$, the PDFs for hard-core bosons and spinless fermions are similar, again plotted there for pairs of eigenstates $\alpha$ and $\beta$ for which $(\mathsf{m}_0)_{\alpha\beta}$ is nonzero.

\begin{figure}[!t]
\begin{center}
\includegraphics[width=0.99\columnwidth]{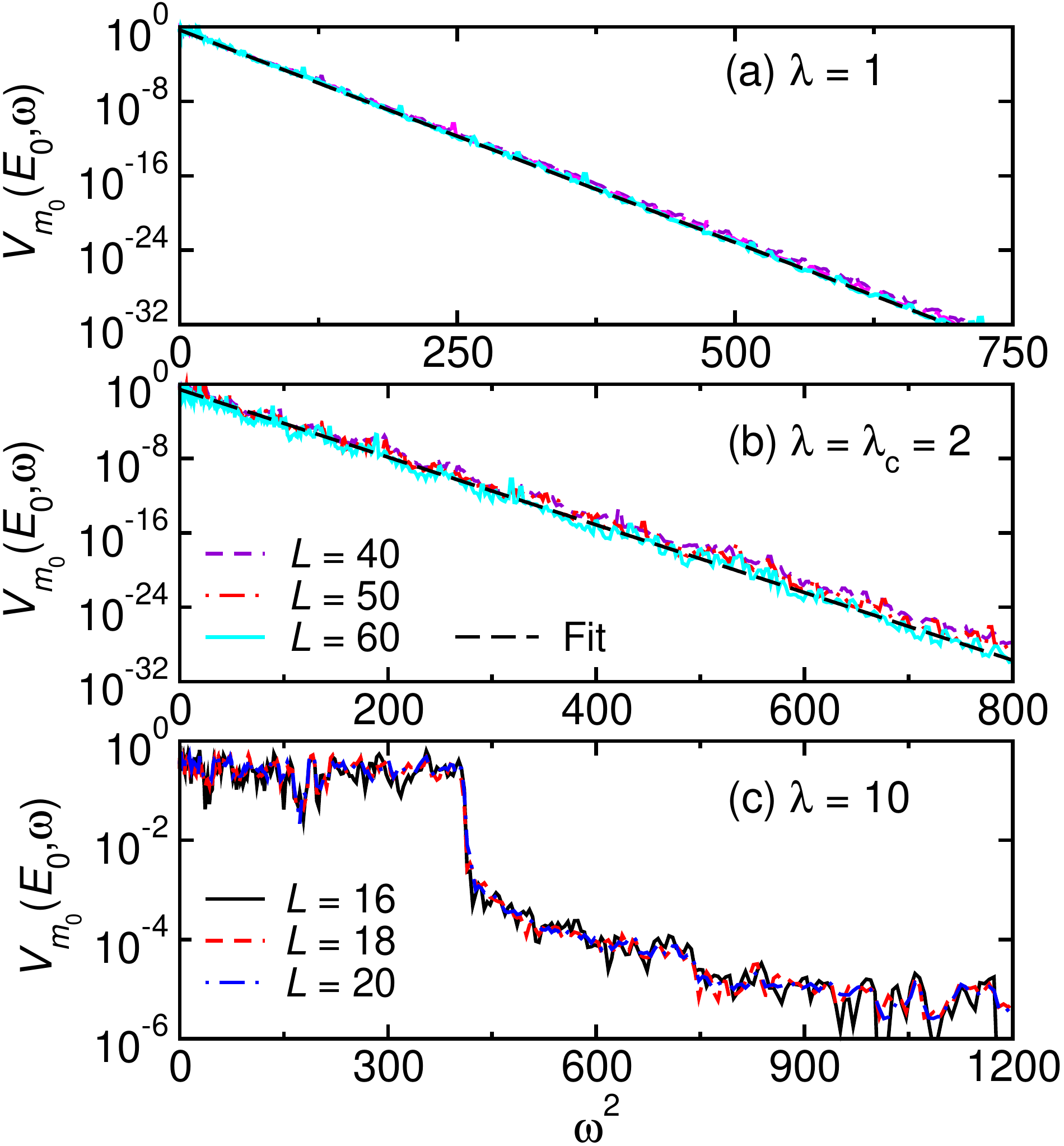}
\caption{\label{fig:Bosonvariance}Scaled variance $V_{m_0}(E_0,\omega)$ of the off-diagonal matrix elements $(m_0)_{\alpha\beta}$ in the hard-core boson Aubry-Andr\'e model plotted vs $\omega^2$. (a) $\lambda=1$ (delocalized regime) and (b) $\lambda=2$ (transition point). For these values of $\lambda$, we show results for systems with sizes $L=40$ (dashed lines), 50 (dashed-dotted lines), and 60 (solid lines). The long dashed lines are Gaussian fits to the $L=60$ results for $\omega^2\in[300,600]$, with a fitting parameter [see Eq.~(\ref{eq:gaussian})] $a=0.11$ in (a) and $a=0.08$ in (b). We randomly sample at least $10^8$ pairs of eigenstates with $\Delta E/L= 2\times 10^{-4}$. The average is carried out over at least 1000 Hamiltonian realizations with randomly selected phases $\phi_0$. (c) $\lambda=10$ (localized regime). Results are shown for systems with sizes $L=16$ (solid line), 18 (dashed line), and 20 (dashed-dotted line). For this value of $\lambda$, we consider all pairs of eigenstates with $\Delta E/L= 2\times 10^{-4}$, and average over 40 Hamiltonian realizations with randomly selected phases $\phi_0$. All calculations are carried out at half filling $N=L/2$, and the results are coarse grained using $\Delta\omega=0.05$.} 
\end{center}
\end{figure}

A complementary understanding of what happens to the off-diagonal matrix elements of the hard-core bosons as $\lambda$ increases can be gained studying for the hard-core bosons the matrix elements that are zero in the spinless fermions model. To quantify their magnitude in the hard-core boson system, we calculate
\begin{equation} \label{def_rzero}
    r_{\rm zero} = \frac{1}{||\hat m_0||^2} \sum_{\alpha,\beta \in {\rm zero}} |(m_0)_{\alpha\beta}|^2 \;,
\end{equation}
where the sum is carried out over pairs of eigenstates for which the corresponding spinless-fermions matrix elements vanish ($\alpha,\beta \in {\rm zero}$, the overwhelming majority of pairs of eigenstates). Results for $r_{\rm zero}$ vs $\lambda$ are shown in the inset of Fig.~\ref{fig:Bosondistribution} for $L=20$. In the delocalized regime, $r_{\rm zero}$ is close to one. This is consistent with the off-diagonal matrix elements being dense. In the localized regime $r_{\rm zero} \to 0$ as $\lambda$ increases, which shows that in this regime the magnitude of those matrix elements decreases as the others (the ``nonzero'' ones) become similar the ones of the fermions.

\begin{figure}[!t]
\begin{center}
\includegraphics[width=0.99\columnwidth]{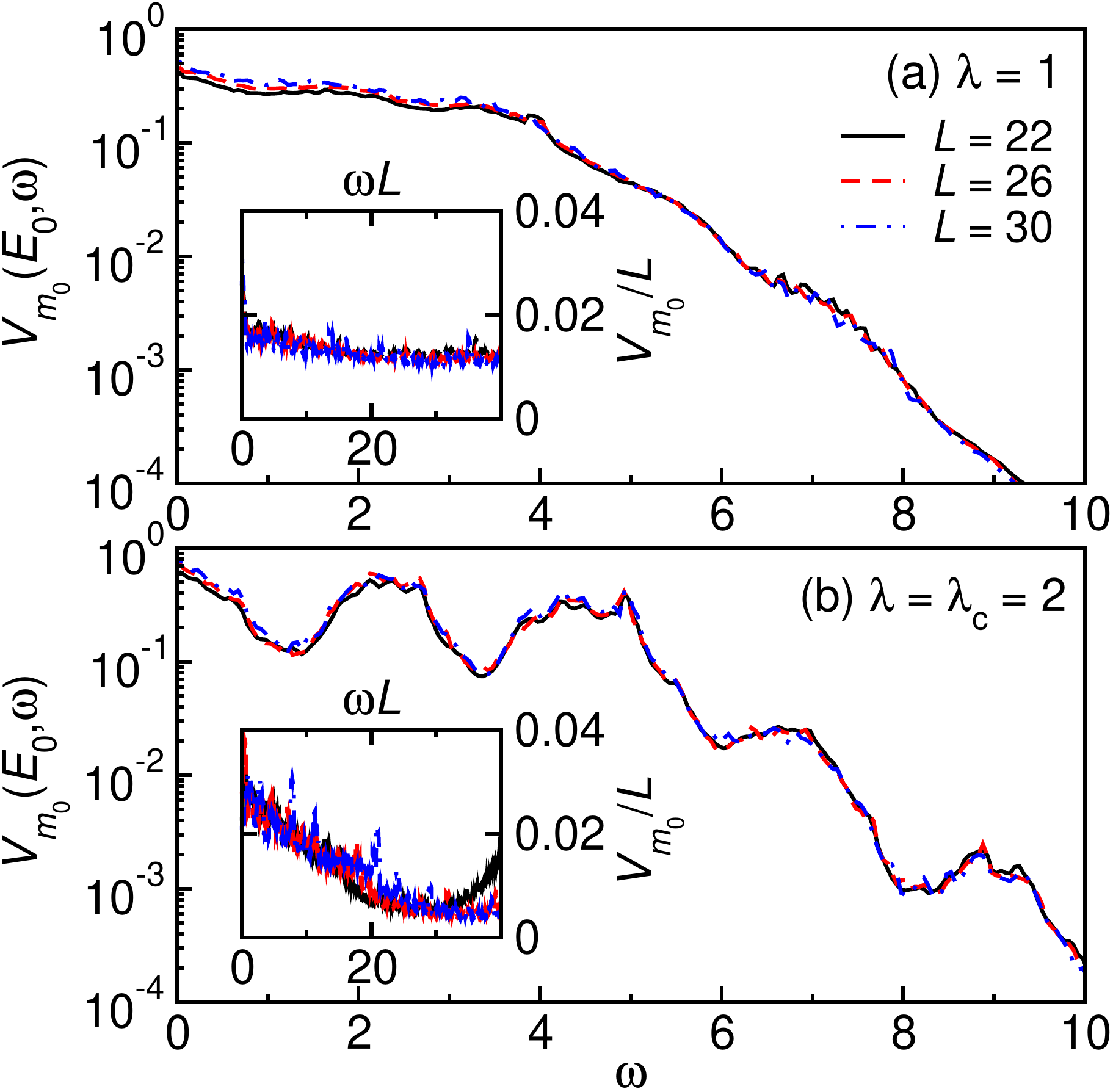}
\caption{\label{fig:AAvariancesmall}$V_{m_0}(E_0,\omega)$ in the hard-core boson Aubry-Andr\'e model at low and intermediate frequencies $\omega$, for (a) $\lambda=1$ and (b) $\lambda=2$. Results are shown for systems with sizes $L=22$ (solid lines), 26 (dashed lines), and 30 (dashed-dotted lines). We randomly select at least $5\times 10^8$ pairs of eigenstates at $\Delta E/L= 2\times 10^{-4}$, and average over at least 5000 Hamiltonian realizations with randomly selected phases $\phi_0$. The variance is coarse grained using a frequency window $\Delta\omega=0.05$. (Insets) The same results as in the main panels but rescaled to show $V_{m_0}(E_0,\omega)/L$ vs $\omega L$. The variance is coarse grained using a frequency window $\Delta\omega=0.02$, and plotted as a running average.} 
\end{center}
\end{figure}

In Fig.~\ref{fig:Bosonvariance}, we show results for the scaled variance $V_{m_0}(E_0,\omega)$ as a function of $\omega^2$ for different system sizes and values of $\lambda$. (In order to reduce finite size effects, in these and in the calculations that follow we carry out an average over results obtained for Aubry-Andr\'e Hamiltonians with randomly selected phases $\phi_0$.) As one may have advanced given the results in Fig.~\ref{fig:Bosondistribution}, the results for the variance are very different in the delocalized and localized regimes. In the delocalized regime [Fig.~\ref{fig:Bosonvariance}(a)] and at the transition point [Fig.~\ref{fig:Bosonvariance}(b)], $V_{m_0}$ exhibits a Gaussian decay at high frequencies [similar to the one observed for translationally invariant hard-core bosons in Fig.~\ref{fig:PWvariance}(b)]. On the other hand, Fig.~\ref{fig:Bosonvariance}(c) shows that no such a Gaussian decay occurs in the localized regime, similar to what happens for noninteracting fermions. For $\lambda=10$ in Fig.~\ref{fig:Bosonvariance}(c), one can see a sort of plateau in the variance for $\omega \lesssim 20$ (the bandwidth of the single-particle spectrum is $\omega \sim 20$). This result is similar to the one for spinless fermions at the same $\lambda=10$ in Fig.~\ref{fig:Fermionvariance}(c). For $\omega \gtrsim 20$ in Fig.~\ref{fig:Bosonvariance}(c), $V_{m_0}$ exhibits a sharp drop. In contrast to the fermions, however, $V_{m_0}(\omega \gtrsim 20)$ for hard-core bosons is small but nonzero.

\begin{figure}[!t]
\begin{center}
\includegraphics[width=0.99\columnwidth]{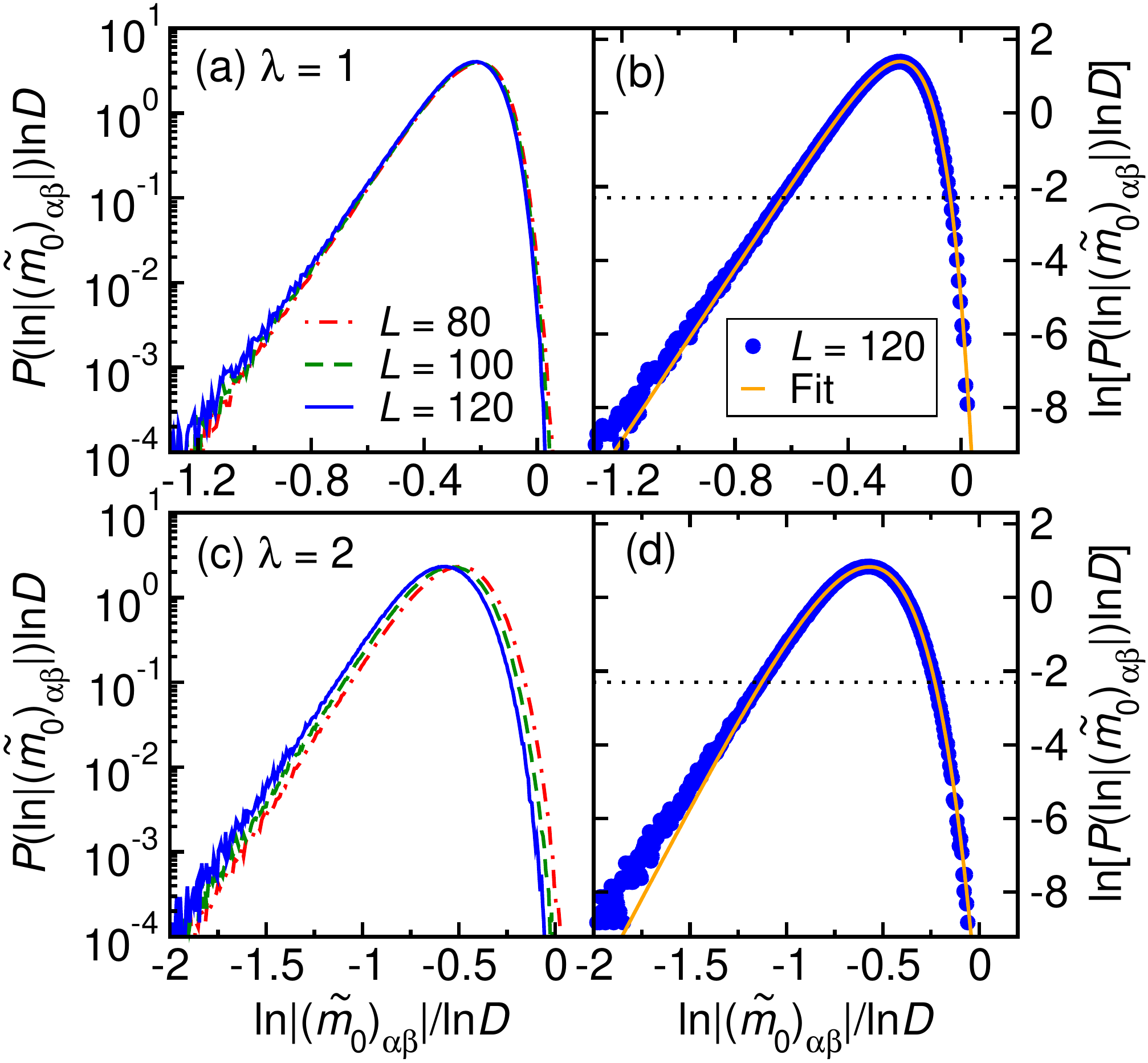}
\caption{\label{fig:AAscale}Scaled probability density function $P(\ln|(\tilde{m}_0)_{\alpha\beta}|) \ln D$ vs $\ln |(\tilde{m}_0)_{\alpha\beta}|/\ln D$ in the hard-core boson Aubry-Andr\'e model. (a, c) Results for $\lambda=1$ and $\lambda=2$, respectively, for systems with sizes $L=80$ (solid lines), 100 (dashed lines), and 120 (dashed-dotted lines) at half filling $N=L/2$. (b, d) The symbols show the results for $L=120$ from panels (a) and (c), respectively. The solid line is a fit to the results above the horizontal dotted line [$P(\ln|(\tilde{m}_0)_{\alpha\beta}|) \ln D \geq 0.1$] using the function in Eq.~(\ref{eq:fit1}). The fitting parameters are: $A_0=5.06$, $B_0=10.08$, $k_0=11.62$, $x_0=-0.23$ for $\lambda=1$, and $A_0=9.40$, $B_0=3.50$, $k_0=10.02$, $x_0=-0.87$ for $\lambda=2$. Pairs of eigenstates are sampled randomly for $\Delta E/L= 2\times 10^{-4}$ and $\omega=7$ with $\Delta\omega=0.05$. We average over at least $3\times 10^6$ pairs of eigenstates, and over at least 600 Hamiltonian realizations for randomly selected phases $\phi_0$.} 
\end{center}
\end{figure}

We emphasize that we use a different numerical protocol in the calculations of the off-diagonal matrix elements in the delocalized regime and at the transition point ($\lambda \leq 2$), compared to the one in the localized regime ($\lambda > 2$). Given the dense nature of the matrix elements in the delocalized regime, for $\lambda \leq 2$ we can carry out calculations for large system sizes randomly sampling matrix elements that belong to the target energy window. On the other hand, in the localized regime for hard-core bosons (as in any regime in the spinless fermion case), the variances is dominated by a vanishingly small fraction of the matrix elements. In those cases, we need to compute all pairs of eigenstates in the target energy window, thereby limiting the calculations to systems with sizes $L \leq 22$ for hard-core bosons. In the reminder of this section we focus on values of $\lambda \leq 2$ because those are the ones for which we expect the properties of hard-core boson matrix elements to resemble those in integrable interacting systems not mappable onto noninteracting models, such as the spin-1/2 XXZ chain.

In Fig.~\ref{fig:AAvariancesmall} we show the scaled variance $V_{m_0}(E_0,\omega)$ vs $\omega$ at low and intermediate frequencies for $\lambda=1$ [Fig.~\ref{fig:AAvariancesmall}(a)] and $\lambda=2$ [Fig.~\ref{fig:AAvariancesmall}(b)]. The results at intermediate frequencies collapse for different system sizes $L$. At low frequencies $\omega\propto 1/L$, the inset shows that the results collapse when plotting $V_{m_0}/L$ vs $\omega L$, as discussed before for the translationally invariant case. Overall, up to additional structure in the variances of the Aubry-Andr\'e case, the results in Fig.~\ref{fig:AAvariancesmall} are qualitatively similar to those reported in Fig.~\ref{fig:PWvariance}(a).

The corresponding scaled PDFs are shown in Figs.~\ref{fig:AAscale}(a) and~\ref{fig:AAscale}(c) for $\lambda=1$ and 2, respectively, for different system sizes and $\omega=7$. As in the translationally invariant case, the curves collapse in the delocalized regime [$\lambda=1$ in Fig.~\ref{fig:AAscale}(a)]. The collapse worsens at the transition point [$\lambda=2$ in Fig.~\ref{fig:AAscale}(c)]. The latter finding suggests that further rescaling may be needed at $\lambda_c$, a point whose detailed investigation is postponed to future studies. In Figs.~\ref{fig:AAscale}(b) and~\ref{fig:AAscale}(d) we show that the scaled PDFs, both for $\lambda=1$ and 2, are well described by the ansatz in Eq.~(\ref{eq:fit1}) with parameters that depend on the Hamiltonian parameters. Hence, the corresponding distributions $P(|(\tilde{m}_0)_{\alpha\beta}|^2)$ are well described by generalized Gamma distributions, see Eq.~\eqref{eq:prob1}.

\section{Beyond hard-core boson models} \label{sec:xxz}

The main goal of this work has been the study of the PDFs of the off-diagonal matrix elements of a specific few-body operator in models of hard-core bosons that can be mapped onto noninteracting spinless fermions (in order to be able to study large system sizes, $L\sim 100$), which we expect to describe the PDFs of the off-diagonal matrix elements of operators in integrable interacting models that are not mappable onto noninteracting models (for which full exact diagonalization studies are limited to sizes $L\sim 20$). The goal of this section is to provide evidence to support our expectation that the main result for the PDFs in the previous sections applies beyond hard-core bosons models. Specifically, we show that the same generalized Gamma distributions that describe the distributions of off-diagonal matrix elements of the occupation of the zero quasimomentum mode of hard-core bosons describe the distribution of off-diagonal matrix elements of a local operator in the spin-1/2 XXZ model~\cite{LeBlond_Rigol_Eigenstate_20}. In Ref.~\cite{LeBlond_Rigol_Eigenstate_20}, the distributions of the off-diagonal matrix elements were reported for $\omega\rightarrow0$. In what follows, we first discuss results for hard-core bosons in the limit $\omega\rightarrow0$ before discussing the results from Ref.~\cite{LeBlond_Rigol_Eigenstate_20}.

\subsection{Hard-core boson distributions for $\omega\rightarrow0$}

In the previous sections, we focused on the distribution of the off-diagonal matrix elements in the intermediate frequency regime ($\omega=7$). We did this in order to avoid the low-frequency ``ballistic'' scaling of the off-diagonal matrix elements with $L$. To study the distribution of the off-diagonal matrix elements for $\omega\rightarrow 0$, we need to consider the following scaled matrix elements 
\begin{equation}\label{def_tilde_m02}
|(\tilde{m}^*_0)_{\alpha\beta}|^2=|(m_0)_{\alpha\beta}\sqrt{D}/\sqrt{L}|^2 \;,
\end{equation}
which have an extra $1/\sqrt{L}$ factor when compared to $|(\tilde{m}_0)_{\alpha\beta}|$ in Eq.~(\ref{def_tilde_m0}). The scaled matrix elements $(\tilde{m}^*_0)_{\alpha\beta}$ are the ones that are $O(1)$ in the thermodynamic limit.

\begin{figure}[!b]
\begin{center}
\includegraphics[width=0.99\columnwidth]{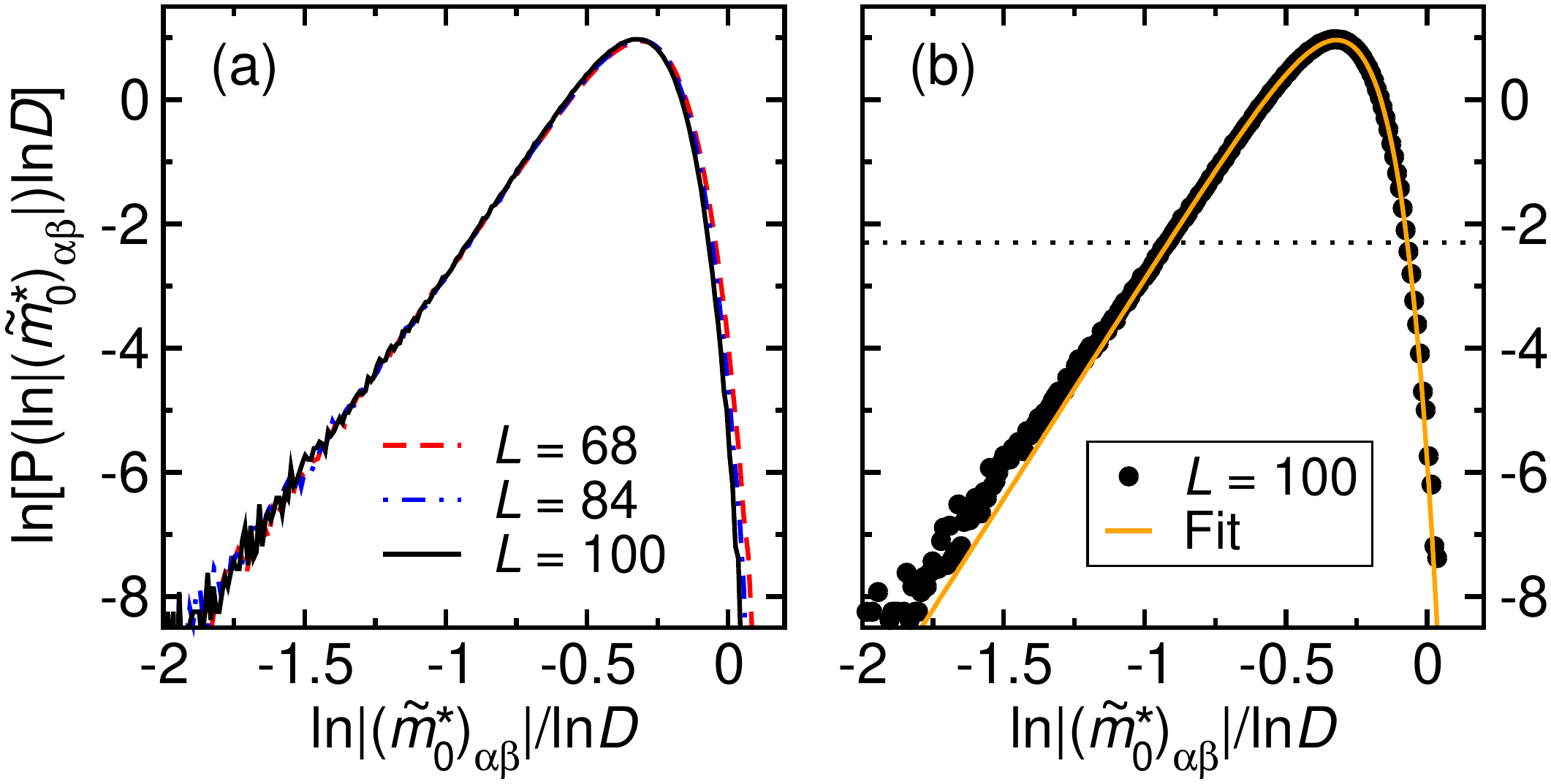}
\caption{\label{fig:omega0} (a) Scaled probability density function $P(\ln|(\tilde{m}^*_0)_{\alpha\beta}|) \ln D$ [see Eq.~(\ref{def_tilde_m02})] vs $\ln |(\tilde{m}^*_0)_{\alpha\beta}|/\ln D$ in the translationally invariant hard-core boson model considered in Sec.~\ref{sec:HCBstranslation}. We study eigenstates in the quasimomentum sector $\kappa = 2\pi/L$ for systems at quarter filling $N = L/4$. We show results for systems with sizes $L=68$ (dashed line), 84 (dashed-dotted line), and 100 (solid line). We randomly select at least $3\times 10^6$ pairs of eigenstates with $\Delta E/L= 2\times 10^{-4}$ and $\omega\in[0,0.05]$. (b) The symbols show the logarithm of the results for $L=100$ in (a), and the solid line is a fit to the points above the dotted line [$P(\ln|(\tilde{m}^*_0)_{\alpha\beta}|) \ln D \geq 0.1$] using the function in Eq.~(\ref{eq:fit1}). The fitting parameters are $A_0=4.22$, $B_0=7.25$, $k_0=7.10$, and $x_0=-0.32$.} 
\end{center}
\end{figure}

In Fig.~\ref{fig:omega0}, we show the scaled probability density function $P(\ln|(\tilde{m}^*_0)_{\alpha\beta}|) \ln D$ as a function of $\ln |(\tilde{m}^*_0)_{\alpha\beta}|/\ln D$ in the translationally invariant hard-core boson model discussed in Sec.~\ref{sec:HCBstranslation}. All the parameters used in the calculations are the same as the ones used for Fig.~\ref{fig:PWscale}, except for the frequency range $\omega\in[0,0.05]$. In Fig.~\ref{fig:omega0}(a), one can see that the curves collapse for different system sizes $L$. In Fig.~\ref{fig:omega0}(b), we fit the scaled PDF with the ansatz function from Eq.~(\ref{eq:fit1}). The outcome of the fitting agrees well with the numerical results, with similar fitting parameters as the ones obtained in Fig.~\ref{fig:PWscale}. Thus, the PDF of $|(\tilde{m}^*_0)_{\alpha\beta}|^2$ is well described by a generalized Gamma distribution,
\begin{equation}\label{eq:prob2}
    P(|(\tilde{m}^*_0)_{\alpha\beta}|^2)=P_D |(\tilde{m}^*_0)_{\alpha\beta}|^{2(k_D-1)}\exp\left[-\alpha_D |(\tilde{m}^*_0)_{\alpha\beta}|^{2B_D}\right]\,,
\end{equation} 
which is nothing but Eq.~(\ref{eq:prob1}) after changing $|(\tilde{m}_0)_{\alpha\beta}|^2\rightarrow|(\tilde{m}^*_0)_{\alpha\beta}|^2$.

An interesting property of $P(|(\tilde{m}^*_0)_{\alpha\beta}|^2)$ is that, for $|(\tilde{m}^*_0)_{\alpha\beta}|^2\rightarrow 0$ in large systems sizes, $P(|(\tilde{m}^*_0)_{\alpha\beta}|^2)\propto 1/(\ln D|(\tilde{m}^*_0)_{\alpha\beta}|^2)\simeq 1/(D|(m_0)_{\alpha\beta}|^2)$, where in the last step we used that $\ln D\simeq L$. A similar result, for a fixed system size, was recently reported in Ref.~\cite{sarang21} for a nonlocal Jordan-Wigner string in the spin-1/2 XX chain. The low-frequency behavior of the matrix elements of integrability breaking perturbations in integrable models can be used to gain an analytic understanding of the system size dependence of the onset of many-body quantum chaos~\cite{sarang21}. Our results for the full PDFs in finite systems sizes, and their scaling with system size, can be used in such calculations to improve our understanding of the onset of quantum chaos.

\subsection{Spin-1/2 XXZ model}

\begin{figure}[!b]
\begin{center}
\includegraphics[width=0.99\columnwidth]{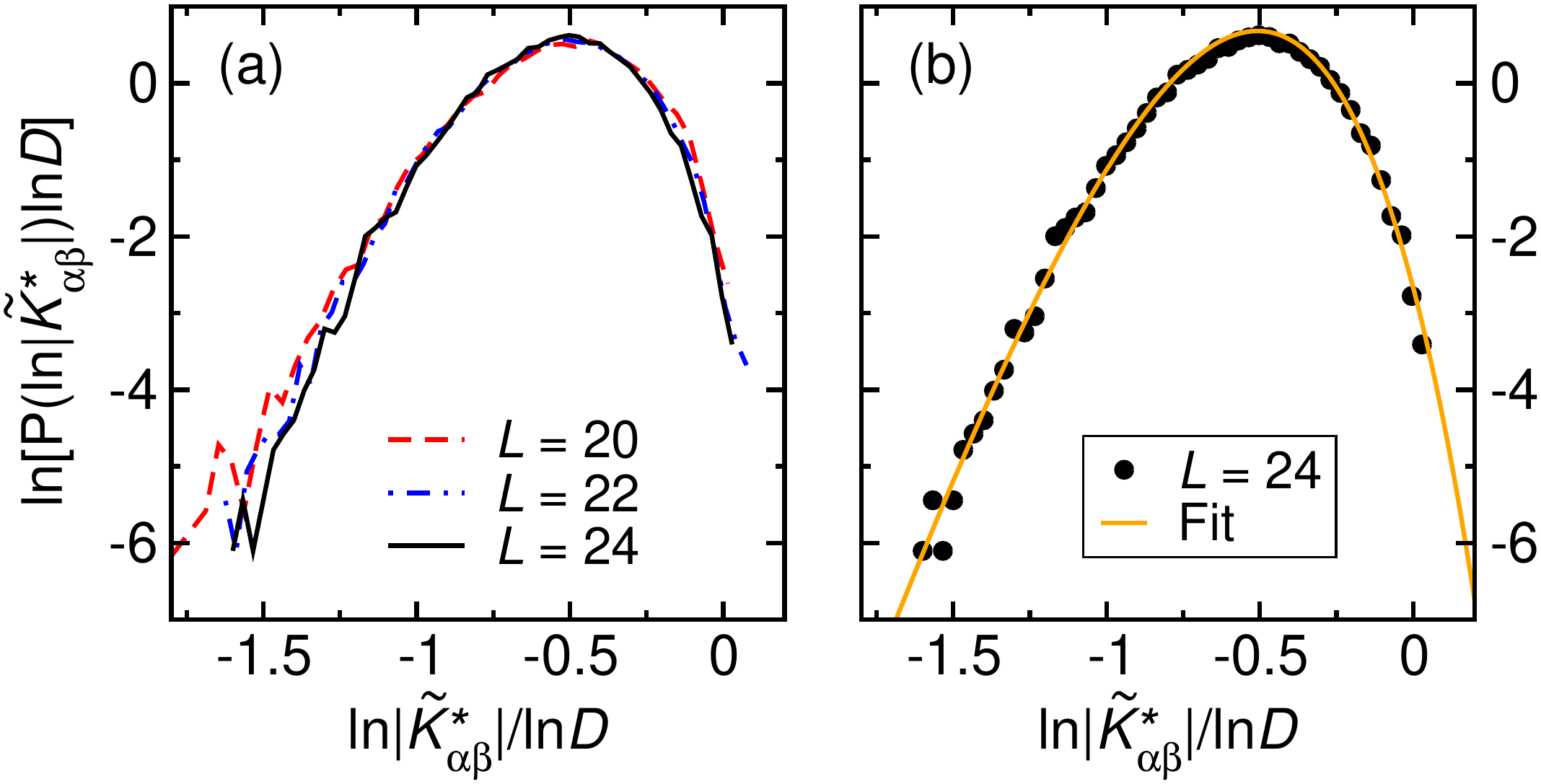}
\caption{\label{fig:XXZ}Scaled probability density function $P(\ln|\tilde{K}^*_{\alpha\beta}|) \ln D$ vs $\ln|\tilde{K}^*_{\alpha\beta}|/\ln D$ [see Eq.~\eqref{eq:Krescale}] in the integrable spin-1/2 XXZ model. (a) $P(\ln|\tilde{K}^*_{\alpha\beta}|) \ln D$ for systems with sizes $L=20$ (dashed line), 22 (dashed-dotted line), and 24 (solid line). (b) The symbols are the results for $\ln[P(\ln|\tilde{K}^*_{\alpha\beta}|) \ln D]$ in the system with $L=24$ from (a), while the solid line is a fit to the results using the function in Eq.~(\ref{eq:fit1}). We use all the data points in the fitting, and obtain the following fitting parameters: $A_0=12.2$, $B_0=1.65$, $k_0=11.6$, and $x_0=-1.56$. The data used for this figure are from Fig.~16 in Ref.~\cite{LeBlond_Rigol_Eigenstate_20}.} 
\end{center}
\end{figure}

With the knowledge gained so far, we are ready to revisit the results in Ref.~\cite{LeBlond_Rigol_Eigenstate_20} for the translationally invariant spin-1/2 XXZ chain, whose Hamiltonian has the form
\begin{equation}
    \hat H_{\rm XXZ}=\sum_{j=1}^{L}\left[\frac{1}{2}(\hat S^+_{j} \hat S^-_{j+1}+{\rm H.c.}) +\Delta \hat S^z_{j}\hat S^z_{j+1}\right]\,,
\end{equation}
where $\hat S^{x\,(y,z)}_i$ are spin-1/2 operators in the $x$ ($y$, $z$) directions on site $j$, $\hat S^\pm_j=\hat S^x_j\pm i\hat S^y_j$ are the corresponding ladder operators. One of the operators studied in Ref.~\cite{LeBlond_Rigol_Eigenstate_20} is the next-nearest-neighbor flip-flop operator
\begin{equation}
    \hat K=\hat S^+_{1} \hat S^-_{3}+\hat S^+_{3} \hat S^-_{1}\,,
\end{equation}
and we are interested in results reported there for the matrix elements of $\hat K$ in pairs of eigenstates within the same quasimomentum sectors, specifically, in the results reported in Fig.~16 of Ref.~\cite{LeBlond_Rigol_Eigenstate_20} for $\Delta=0.55$, where the pairs of eigenstates were taken to have an average energy $|\bar E|\leq 0.025L$, and 40\,000 off-diagonal matrix elements were selected that correspond to the lowest values of $\omega$. Following Eq.~\eqref{def_tilde_m02}, we study the PDF of the scaled matrix elements
\begin{equation} \label{eq:Krescale}
    |\tilde K^*_{\alpha\beta}|^2 = |K_{\alpha\beta}\sqrt{D}/\sqrt{L}|^2 \;,
\end{equation}
when reanalyzing the data from Ref.~\cite{LeBlond_Rigol_Eigenstate_20}.

In Fig.~\ref{fig:XXZ}(a), we replot the data in Fig.~16 of Ref.~\cite{LeBlond_Rigol_Eigenstate_20} using the additional rescalings in Eqs.~\eqref{eq:rescm0} and~\eqref{def_Pln_rescaling}. The results for different system sizes in Fig.~\ref{fig:XXZ}(a) exhibit a good collapse (note that the system sizes are much smaller than those in Figs.~\ref{fig:PWscale} and~\ref{fig:AAscale}). In Fig.~\ref{fig:XXZ}(b) we compare the results for the largest system size to a fit to the ansatz in Eq.~(\ref{eq:fit1}). There is also a good agreement between the numerical results (symbols) and the fit (solid line). This suggests that the off-diagonal matrix elements of observables in integrable interacting models are generically described by generalized Gamma distributions.

\section{Summary} \label{sec:summary}

We studied the statistical properties of the matrix elements of few-body operators in hard-core boson models, and of noninteracting spinless fermions to which hard-core bosons can be mapped, in one-dimensional lattices. We showed, first analytically and then in numerical calculations of the model of interest, that the off-diagonal matrix elements of few-body operators in the eigenstates of noninteracting fermionic Hamiltonians are {\it sparse}, i.e, the overwhelming majority of the matrix elements vanishes (the number of nonzero matrix elements scales polynomially with the system size). For hard-core bosons on the other hand, we showed that there are few-body operators, such as the occupation of quasimomentum modes that are of experimental relevance, for which the off-diagonal matrix elements are {\it dense}, i.e., the overwhelming majority of the matrix elements are nonzero.

We considered two hard-core boson Hamiltonians that can be mapped onto noninteracting spinless fermions Hamiltonians, translationally invariant hard-core bosons with nearest neighbor hoppings [Eq.~\eqref{eq:Hhcb_pw}] and the hard-core boson Aubry-Andr\'e model [Eq.~\eqref{eq:HhcbAA}]. For translationally invariant hard-core bosons and for the hard-core boson Aubry-Andr\'e model in the delocalized regime, we showed that the scaled variances of the off-diagonal matrix elements of the occupation of the zero quasimomentum mode behave as those of local operators in integrable interacting models that are not mappable onto noninteracting models, such as the spin-1/2 XXZ chain~\cite{LeBlond_Mallayya_19, Brenes_LeBlond_20, brenes_goold_20, LeBlond_Rigol_Eigenstate_20}. Namely, they exhibit a regime with a Gaussian decay in $\omega$ at high $\omega$, and a regime with a ballistic scaling when $\omega\propto1/L$. On the other hand, we found the behavior of the off-diagonal matrix elements to be completely different in the localized regime, in which the variance is strongly suppressed at frequencies beyond the single-particle bandwidth and no Gaussian decay occurs at high frequencies. The off-diagonal matrix elements also become sparse as $\lambda$ increases in that regime, and become similar to those of the noninteracting spinless fermions to which hard-core bosons can be mapped.

Our main results in this work are first the rescaling of the off-diagonal matrix elements of hard-core bosons in delocalized regimes, involving the logarithm of the Hilbert space dimension [see Eq.~\eqref{eq:rescm0}], and the corresponding rescaling of the PDFs [see Eq.~\eqref{def_Pln_rescaling}], to produce meaningful PDFs in the thermodynamic limit. The second main result is the finding that the PDFs after rescaling are well described by generalized Gamma distributions [see Eq.~\eqref{eq:prob1}]. Studying translationally invariant hard-core bosons and the hard-core boson Aubry-Andr\'e model we showed that these distributions are robust against the breaking of translational symmetry, so long as the system does not localize. We also found that the values of the parameters in the generalized Gamma distributions depend on the Hamiltonian considered and its parameters.

Furthermore, a reanalysis of the results for the translationally invariant spin-1/2 XXZ chain in Ref.~\cite{LeBlond_Rigol_Eigenstate_20} suggests that generalized Gamma distributions generically describe the PDFs of the off-diagonal matrix elements of observables in integrable interacting models. Further studies are needed to support this conjecture and to understand why such distributions describe the off-diagonal matrix elements of observables in integrable systems. Well known distributions that are special cases of the generalized Gamma distribution in Eq.~\eqref{eq:prob1} include the Weibull distribution ($k^*_D=B^*_D$), the Gamma distribution ($B^*_D=1$), and the exponential distribution ($k^*_D=B^*_D=1$). 

\section{Acknowledgments}
This work was supported by the National Science Foundation, Grant No.~2012145 (Y.Z. and M.R.), and by the Slovenian Research Agency (ARRS), Research core fundings Grants No.~P1-0044 and No.~J1-1696 (L.V.). We grateful to Sarang Gopalakrishnan for stimulating discussions.

\appendix

\section{Effect of random sampling} \label{app:sampling}

Throughout the main text, we have shown results for hard-core boson models that were obtained after calculating all the matrix elements in the target energy and frequency windows in small system sizes, as well as results obtained after randomly sampling the matrix elements in the target energy and frequency windows in larger system sizes. Here compare both approaches in small systems sizes.

\begin{figure}[!t]
\begin{center}
\includegraphics[width=0.99\columnwidth]{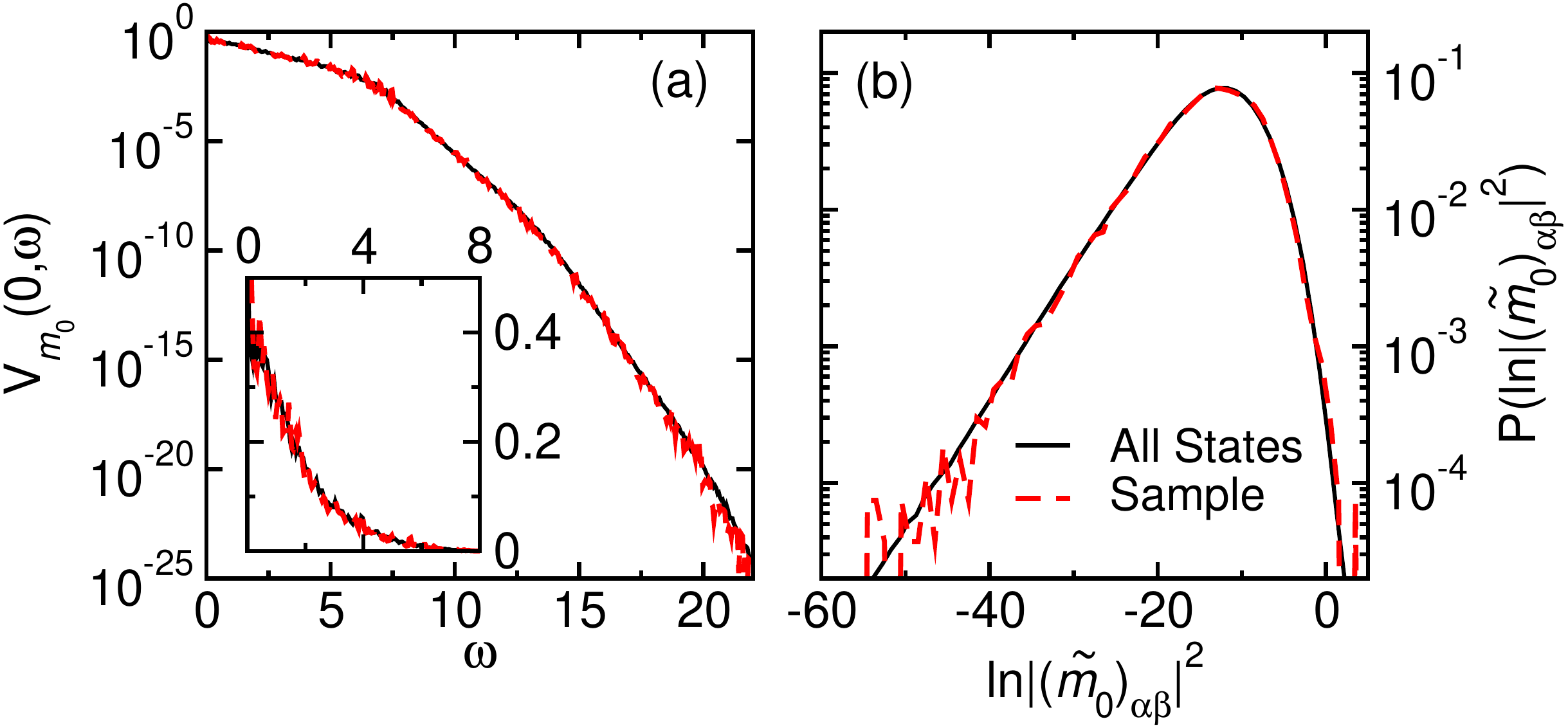}
\caption{\label{fig:sample}(a) Effect of random sampling when calculating the scaled variance $V_{m_0}(0,\omega)$. We show results for translationally invariant hard-core bosons in a system with $L=36$ at quarter filling $N=9$, and consider energy eigenstates with total quasimomentum $\kappa=2\pi/L$, and $\Delta E/L=2\times 10^{-4}$. The solid lines show the results when all pairs of eigenstates are included (a total of $7.3\times 10^9$ pairs), while the dashed lines show the results obtained using $4\times10^6$ randomly sampled pairs of eigenstates. (Inset) The same results as in the main panels but with $V_{m_0}(0,\omega)$ plotted in a linear scale. (b) Effect of random sampling for the probability density function $P(\ln |(\tilde{m}_0)_{\alpha\beta}|^2)$ at $\omega=7$ ($\Delta \omega=0.1$). The data used for this plot are the same data used in (a).}
\end{center}
\end{figure}

In Fig.~\ref{fig:sample}(a) and its inset we show the scaled variance $V_{m_0}(E_0,\omega)$ from Eq.~(\ref{eq:scaledvar}) in the translationally invariant model. We consider a system with $L=36$ sites and $N=9$ particles, and focus on pairs of eigenstates from the $\kappa=2\pi/L$ total quasimomentum sector. The black solid lines are the results obtained including all possible pairs of eigenstates in the calculation (a total of $7.3\times 10^9$ pairs), while the red dashed lines are the results obtained using $4\times10^6$ randomly sampled pairs of eigenstates. In both the main panel and its inset we observe a good agreement between the results including all the eigenstates and the results with the random sampling of eigenstate pairs.

In Fig.~\ref{fig:sample}(b) we show the corresponding results for the probability density function $P(|(\tilde{m}_0)_{\alpha\beta}|^2)$ at $\omega=7$ (with $\Delta \omega=0.1$). The total number of eligible pairs of states in this frequency window is around $5.0\times 10^7$, and the sampled data set has $2.7\times10^4$ pairs. Also in this case the agreement between averaging over all pairs of eigenstates and sampling them is excellent. 

More generally, we verified that all the plots shown in the paper do not change visibly if we use one half of the randomly selected pairs of eigenstates in the calculations. On the other hand, in the localized regime of the hard-core boson Aubry-Andr\'e model and in the spinless fermion models, only a small fraction of eigenstate pairs (that vanishes in thermodynamic limit) contributes significantly to the quantities computed in this work. Consequently, in these cases we report results only when all the pairs of matrix elements are computed.\\

\section{PDFs of $\hat m_{k=\pi/2}$ and $\hat m_{k=\pi}$} \label{app:mkpi}

\begin{figure}[!h]
\begin{center}
\includegraphics[width=0.99\columnwidth]{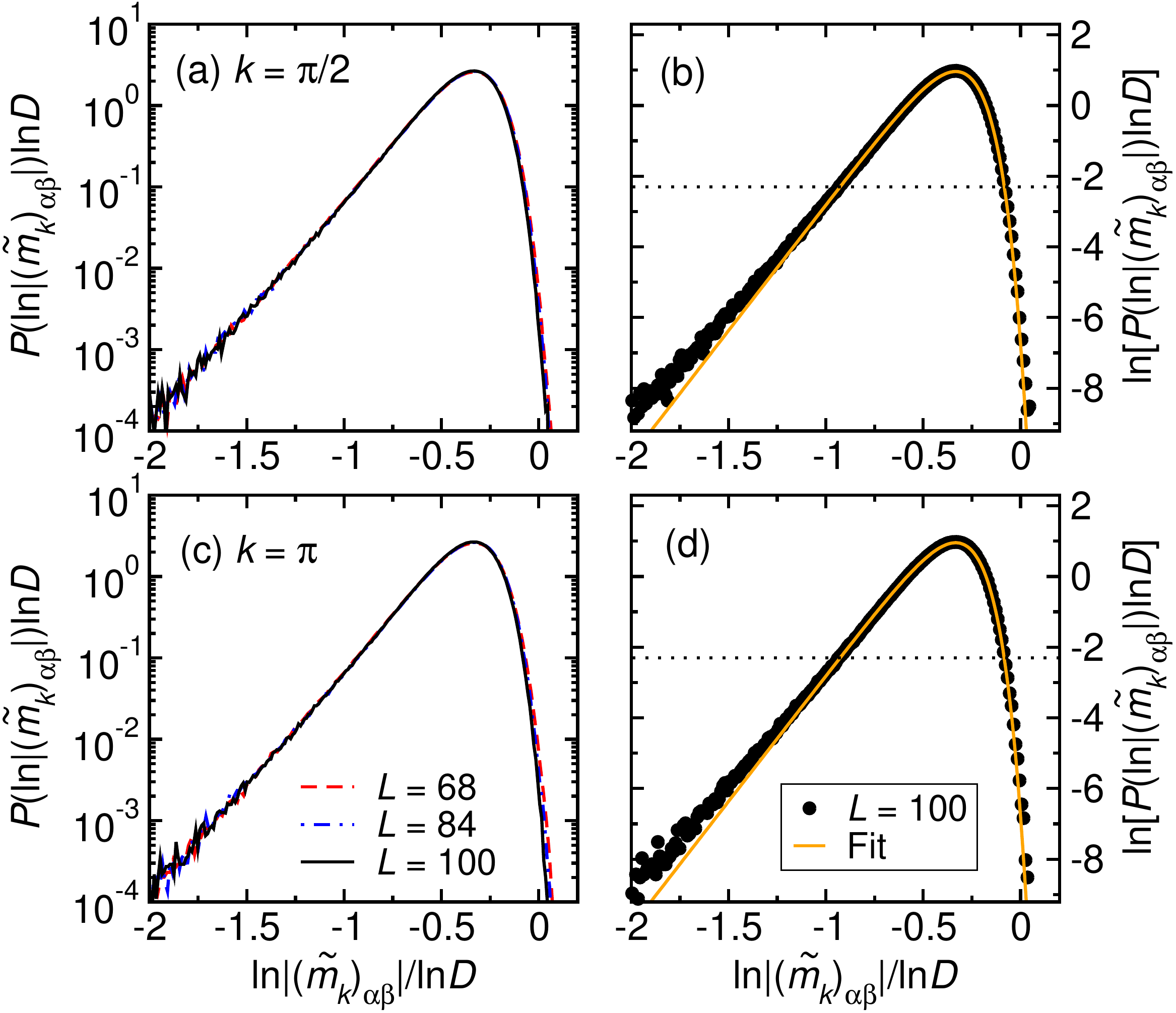}
\caption{\label{fig:otherk}(a, b) Same as Fig.~\ref{fig:PWscale}(a) and~\ref{fig:PWscale}(b), respectively, for $\hat m_{k=\pi/2}$. The fitting parameters in (b) are: $A_0=4.31$, $B_0=7.28$, $k_0=7.12$, and $x_0=-0.33$. (c, d) Same as Fig.~\ref{fig:PWscale}(a) and~\ref{fig:PWscale}(b), respectively, for $\hat m_{k=\pi}$. The fitting parameters in (d) are: $A_0=4.30$, $B_0=7.29$, $k_0=7.11$, and $x_0=-0.33$.}
\end{center}
\end{figure}

In the main text we showed only results for the occupation of the zero quasimomentum mode $\hat m_{0} \equiv \hat m_{k=0}$ [see Eq.~(\ref{eq:mkhcb})]. In Fig.~\ref{fig:otherk}, we show that the PDFs of the off-diagonal matrix elements of $\hat m_{k=\pi/2}$ and $\hat m_{k=\pi}$ are qualitatively (and quantitatively) similar to those for $\hat m_{k=0}$ in the translationally invariant model. Qualitatively similar results (not shown here) were obtained for the Aubry-Andr\'e model in the delocalized regime.

\bibliographystyle{biblev1.bst}
\bibliography{references}

\end{document}